\documentclass[12pt]{article}
\usepackage{graphicx}
\usepackage{lscape}
\usepackage{citesort}
\usepackage{amssymb}
\usepackage{appendix}
\usepackage{multirow}

\begin{document}
\def\qq{\langle \bar q q \rangle}
\def\uu{\langle \bar u u \rangle}
\def\dd{\langle \bar d d \rangle}
\def\sp{\langle \bar s s \rangle}
\def\GG{\langle g_s^2 G^2 \rangle}
\def\Tr{\mbox{Tr}}
\def\figt#1#2#3{
        \begin{figure}
        $\left. \right.$
        \vspace*{-2cm}
        \begin{center}
        \includegraphics[width=10cm]{#1}
        \end{center}
        \vspace*{-0.2cm}
        \caption{#3}
        \label{#2}
        \end{figure}
	}
	
\def\figb#1#2#3{
        \begin{figure}
        $\left. \right.$
        \vspace*{-1cm}
        \begin{center}
        \includegraphics[width=10cm]{#1}
        \end{center}
        \vspace*{-0.2cm}
        \caption{#3}
        \label{#2}
        \end{figure}
                }

\def\ds{\displaystyle}
\def\beq{\begin{equation}}
\def\eeq{\end{equation}}
\def\bea{\begin{eqnarray}}
\def\eea{\end{eqnarray}}
\def\beeq{\begin{eqnarray}}
\def\eeeq{\end{eqnarray}}
\def\ve{\vert}
\def\vel{\left|}
\def\ver{\right|}
\def\nnb{\nonumber}
\def\ga{\left(}
\def\dr{\right)}
\def\aga{\left\{}
\def\adr{\right\}}
\def\lla{\left<}
\def\rra{\right>}
\def\rar{\rightarrow}
\def\lrar{\leftrightarrow}  
\def\nnb{\nonumber}
\def\la{\langle}
\def\ra{\rangle}
\def\ba{\begin{array}}
\def\ea{\end{array}}
\def\tr{\mbox{Tr}}
\def\ssp{{\Sigma^{*+}}}
\def\sso{{\Sigma^{*0}}}
\def\ssm{{\Sigma^{*-}}}
\def\xis0{{\Xi^{*0}}}
\def\xism{{\Xi^{*-}}}
\def\qs{\la \bar s s \ra}
\def\qu{\la \bar u u \ra}
\def\qd{\la \bar d d \ra}
\def\qq{\la \bar q q \ra}
\def\gGgG{\la g^2 G^2 \ra}
\def\q{\gamma_5 \not\!q}
\def\x{\gamma_5 \not\!x}
\def\g5{\gamma_5}
\def\sb{S_Q^{cf}}
\def\sd{S_d^{be}}
\def\su{S_u^{ad}}
\def\sbp{{S}_Q^{'cf}}
\def\sdp{{S}_d^{'be}}
\def\sup{{S}_u^{'ad}}
\def\ssp{{S}_s^{'??}}

\def\sig{\sigma_{\mu \nu} \gamma_5 p^\mu q^\nu}
\def\fo{f_0(\frac{s_0}{M^2})}
\def\ffi{f_1(\frac{s_0}{M^2})}
\def\fii{f_2(\frac{s_0}{M^2})}
\def\O{{\cal O}}
\def\sl{{\Sigma^0 \Lambda}}
\def\es{\!\!\! &=& \!\!\!}
\def\ap{\!\!\! &\approx& \!\!\!}
\def\ar{&+& \!\!\!}
\def\ek{&-& \!\!\!}
\def\kek{\!\!\!&-& \!\!\!}
\def\cp{&\times& \!\!\!}
\def\se{\!\!\! &\simeq& \!\!\!}
\def\eqv{&\equiv& \!\!\!}
\def\kpm{&\pm& \!\!\!}
\def\kmp{&\mp& \!\!\!}
\def\mcdot{\!\cdot\!}
\def\erar{&\rightarrow&}

% .........................................................

\def\simlt{\stackrel{<}{{}_\sim}}
\def\simgt{\stackrel{>}{{}_\sim}}

% .........................................................

\renewcommand{\textfraction}{0.2}    %float (figures) parameters
\renewcommand{\topfraction}{0.8}   

\renewcommand{\bottomfraction}{0.4}   
\renewcommand{\floatpagefraction}{0.8}
\newcommand\mysection{\setcounter{equation}{0}\section}

\def\baeq{\begin{appeq}}     \def\eaeq{\end{appeq}}  
\def\baeeq{\begin{appeeq}}   \def\eaeeq{\end{appeeq}}
\newenvironment{appeq}{\beq}{\eeq}   
\newenvironment{appeeq}{\beeq}{\eeeq}
\def\bAPP#1#2{
 \markright{APPENDIX #1}
 \addcontentsline{toc}{section}{Appendix #1: #2}
 \medskip
 \medskip
 \begin{center}      {\bf\LARGE Appendix #1 :}{\quad\Large\bf #2}
% \begin{center}      {\bf\LARGE Appendix  :}{\quad\Large\bf #2}
\end{center}
 \renewcommand{\thesection}{#1.\arabic{section}}
\setcounter{equation}{0}
        \renewcommand{\thehran}{#1.\arabic{hran}}
\renewenvironment{appeq}
  {  \renewcommand{\theequation}{#1.\arabic{equation}}
     \beq
  }{\eeq}
\renewenvironment{appeeq}
  {  \renewcommand{\theequation}{#1.\arabic{equation}}
     \beeq
  }{\eeeq}
\nopagebreak \noindent}

\def\eAPP{\renewcommand{\thehran}{\thesection.\arabic{hran}}}

\renewcommand{\theequation}{\arabic{equation}}
\newcounter{hran}
\renewcommand{\thehran}{\thesection.\arabic{hran}}

\def\bmini{\setcounter{hran}{\value{equation}}
\refstepcounter{hran}\setcounter{equation}{0}
\renewcommand{\theequation}{\thehran\alph{equation}}\begin{eqnarray}}
\def\bminiG#1{\setcounter{hran}{\value{equation}}
\refstepcounter{hran}\setcounter{equation}{-1}
\renewcommand{\theequation}{\thehran\alph{equation}}
\refstepcounter{equation}\label{#1}\begin{eqnarray}}

%       the stuff below defines \eqalign and \eqalignno in such a
%       way that they will run on Latex

\newskip\humongous \humongous=0pt plus 1000pt minus 1000pt
\def\caja{\mathsurround=0pt}
%\def\eqalign#1{\,\vcenter{\openup1\jot
%\caja   %\ialign{\strut \hfil$\displaystyle{##}$&$
%\displaystyle{{}##}$\hfil\crcr#1\crcr}
%}\,}

% ...........................................................

\title{
         {\Large
                 {\bf
Vector meson--baryon strong coupling contants in light cone
QCD sum rules
                 }
         }
      }

\author{\vspace{1cm}\\
{\small T. M. Aliev$^a$ \thanks
{e-mail: taliev@metu.edu.tr}~\footnote{permanent address:Institute
of Physics,Baku,Azerbaijan}\,\,,
A. \"{O}zpineci$^a$ \thanks
{e-mail: ozpineci@p409a.physics.metu.edu.tr}\,\,,
M. Savc{\i}$^a$ \thanks
{e-mail: savci@metu.edu.tr} \,\,,
V. S. Zamiralov$^b$ \thanks  
{e-mail: zamir@depni.sinp.msu.ru}} \\
{\small (a) Physics Department, Middle East Technical University,
06531 Ankara, Turkey} \\
{\small (b) Institute of Nuclear Physics, M. V. Lomonosov MSU, Moscow,
Russia} }
\date{}

\begin{titlepage}
\maketitle
\thispagestyle{empty}

\begin{abstract}
Using the most general form of the interpolating current of the baryons, 
the strong coupling constants of the light vector mesons with the octet 
baryons are calculated within the light cone QCD sum rules. The $SU(3)_f$ 
symmetry breaking effects are taken into account in the calculations. It 
is shown that each of the electric and magnetic coupling constants can be 
described in terms of three universal functions. A detailed comparison of
the results of this work on aforementioned couplings with the existing 
theoretical results is presented.
\end{abstract}

%\vspace{1cm}
~~~PACS number(s): 11.55.Hx, 13.40.Em, 14.20.Jn
\end{titlepage}

\section{Introduction}
   
The strong coupling constants of the pseudoscalar (scalar) octet mesons 
$\pi$, $K$, $\eta$ ($\sigma$, $a_0$, $f_0$), and vector nonet mesons 
$\rho$, $\phi$, $\omega$, $K^\ast$ with baryons are the fundamental
parameters in analysis of the existing experimental results on the 
meson--nucleon, nucleon--hyperon and hyperon--hyperon interactions. The
coupling constants of the vector mesons with the octet baryons can be
written in terms of the $\rho NN$ coupling constant and $\alpha_e$
($\alpha_m$), where $\alpha_e$ ($\alpha_m$) is the $F/(F+D)$ ratio of the
electric (magnetic) coupling constants \cite{Roc01}. The vector dominance
model predicts $\alpha_e = 1$, assuming universal coupling of
the $\rho$ meson to the isospin current \cite{Roc02}. Therefore, reliable
determination of the meson--baryon coupling constants present
an important problem. 
Calculation of these coupling constants from the fundamental theory of
strong interactions, namely QCD, represents a very important task. At the
hadronic scale QCD is nonperturbative, which makes it impossible to
calculate the properties of hadrons from a fundamental QCD Lagrangian.
For this reason, calculation of the properties of hadrons require
nonperturbative methods. Among a number of approaches, especially QCD sum 
rules is one of the most powerful and predictive method \cite{Roc03}.

In this work we calculate the strong coupling constants of the octet vector
mesons with baryons in the framework of the light cone QCD sum rules (LCSR)
method. Note that the $\rho NN$ strong coupling constant is studied in this 
framework in \cite{Roc04}. The strong coupling constants of $\rho NN$,
$\rho\Sigma\Sigma$ and $\rho\Xi\Xi$ are studied in LCSR in \cite{Roc05}.
The coupling constant of the vector mesons 
$\rho$ and $\omega$ with the baryons is studied in the framework of the
external field QCD sum rules method in \cite{Roc06}. The coupling constants 
of pseudoscalar mesons with baryons are studied comprehensively in the framework 
of the light cone version of the QCD sum rules \cite{Roc07}.

Few words about the light cone QCD sum rules (LCSR) method are in order.
This method is based on operator product expansion, which is carried out
over twist near the light cone $x^2\approx 0$. The matrix elements of
nonlocal operators between one particle and vacuum states are parametrized
in terms of distribution amplitudes, which are the main nonperturbative
parameters. More about the LCSR method and its applications, can be found in 
\cite{Roc08,Roc09}. 

The paper is organized as follows. In section 2, $SU(3)_f$ classification of
the vector meson--baryon coupling constants are presented, and they are
calculated in LCSR framework in section 3. Section 4 is dedicated to the
numerical analysis of the sum rules for the above--mentioned coupling
constants and our discussions and comments on these results. In this section
we also present a comparison of our results with the predictions of 
other approaches. 

\section{$SU(3)_f$ classification of the vector meson baryon coupling constants}

It is well known that, in $SU(3)_f$ symmetry, coupling constants of all 
pseudoscalar mesons with baryons can be expressed in terms of two constants
${\cal F}$ and ${\cal D}$ in the following way:
\bea
\label{eoc01}
{\cal L}_{BBM} =  \sqrt{2} {\cal F} \, \mbox{Tr} \bar{B} [V,B] + \sqrt{2} {\cal D}
\, \mbox{Tr} \bar{B} \{V,B\} - {1\over \sqrt{2}} ({\cal F}+{\cal D})
\mbox{Tr} (\bar{B} B) \mbox{Tr}V ~,
\eea
and we assume ideal mixing of the octet and singlet
isosinglets giving observable $\rho^0$ and $\omega$ mesons. Ideal mixing
corresponds to the mixing angle $\theta = \cos^{-1}\sqrt{2/3} = 35.3^0$,
which is very close to the experimental value $\theta = 37.5^0$ \cite{Roc10}. 
The coefficient of the last term is chosen to eliminate the coupling of the
nucleon to the pure $\bar{s}s$ state $\phi$. $B$ and $V$ are the octet
baryons and octet vector mesons: 
\bea
\label{eoc02}
B_\beta^\alpha = \left( \begin{array}{ccc}
{1\over \sqrt{2}} \Sigma^0 + {1\over \sqrt{6}} \Lambda & \Sigma^+ & p \\
\Sigma^- & - {1\over 2} \Sigma^0 + {1\over \sqrt{6}} \Lambda  & n \\
\Xi^- & \Xi^0  & -{2\over \sqrt{6}} \Lambda
\end{array} \right)~,
\eea 
\bea
\label{eoc03}
V_\beta^\alpha = \left( \begin{array}{ccc}
{1\over \sqrt{2}} \rho^0 + {1\over \sqrt{2}} \omega & \rho+ & K^{\ast +} \\
\rho^- & - {1\over \sqrt{2}} \rho^0 + {1\over \sqrt{2}} \omega & K^{\ast 0} \\
K^{\ast -} & \bar{K}^{\ast 0}  & \phi
\end{array} \right)~. 
\eea

Instead of the ${\cal F}$ and ${\cal D}$ as independent parameters, one can
also choose to work with the coupling $g^{p\to p \rho^0}$ and the ratio
$\alpha = {\cal F}/({\cal F}+{\cal D})$. In terms of these parameters,
${\cal F} = \alpha g^{p\to p \rho^0}$, ${\cal D} = (1-\alpha) g^{p\to p
\rho^0}$. Note also that, There are two pairs of ${\cal F}$ and ${\cal D}$
values; one for the electric type and one for the magnetic type couplings.

\section{Light cone sum rules for the vector meson--baryon coupling
constants}

To construct LCSR for the vector 
meson--baryon strong coupling constants, the following correlation function
is considered:
\bea
\label{eoc08}
\Pi^{B_1 \rar B_2 V} = i \int d^4x e^{ipx} \lla V(q) \vel {\cal T} \left\{ 
\eta_{B_2} (x) \bar{\eta}_{B_1} (0) \right\} \ver 0 \rra~,
\eea
where $B_1~(B_2)$ is the initial (final) baryon, $V$ is a vector meson,
$\eta_B$ is the interpolating
current of the corresponding baryon, $q$ is the momentum of $V$ meson,
and ${\cal T}$ is the time ordering product.
The correlation function can be calculated in terms the hadrons, 
as well as in the deep Euclidean region $p^2 \rar - \infty$, 
in terms of the quark and gluon degrees of freedom. 
Using the operator product expansion (OPE) the corresponding sum rules 
are obtained by equating both representations through the dispersion 
relations.

Let us firstly construct the phenomenological part of the correlation
function. For this aim we will insert a complete set of intermediate states 
with the same quantum numbers as the current operators $\eta_B$. After isolating 
the ground state baryons, we get
\bea
\label{eoc09}
\Pi^{B_1 \rar B_2 V}(p_1^2,p_2^2) =  {\lla 0 \vel \eta_{B_2} \ver
B_2(p_2) \rra \over p_2^2-m_2^2} \lla B_2(p_2) V(q) \vel \right.
B_1(p_1) \rra {\lla B_1(p_1) \vel \bar{\eta}_{B_1} \ver        
0 \rra \over p_1^2-m_1^2} + \cdots~, 
\eea
where $p_1 = p_2+q$, $m_i$ is the mass of baryon $B_i$, and $\cdots$
represents the contributions of the higher states and the continuum.

The matrix elements entering Eq. (\ref{eoc09}) are defined as follows:
\bea
\label{eoc10}
\lla 0 \vel \eta_{B_i} \ver B_i(p_i) \rra \es \lambda_{B_i} u(p_i) ~, \\
\label{eoc11}
\lla B_2(p_2) V(q) \vel \right. B_1(p_1) \rra \es \bar{u}(p_2) \Bigg[ f_1
\gamma_\mu - f_2 {i\over m_1 + m_2 } \sigma_{\mu\nu} q^\nu \Bigg]
u(p_1) \varepsilon^\mu~,
\eea
where $\lambda_{B_i}$ is the overlap amplitude for the baryon $B_i$, 
$q^\nu$ is the vector meson four--momentum, $u$ is the Dirac spinor for 
the baryon which is normalized as $\bar{u}u=2m$.

Using Eqs. (\ref{eoc10}) and (\ref{eoc11}), we obtain the following result
for the phenomenological part of the correlation function:
\bea
\label{eoc12}
\Pi^{B_1 \rar B_2} \es { \lambda_{B_1} \lambda_{B_2}\over (p_1^2-m_1^2)
(p_2^2-m_2^2)} \varepsilon^\mu (\rlap/p_2 + m_2) \Bigg\{ f_1 \gamma_\mu -
f_2 {i \over m_1 + m_2} \sigma_{\mu\nu} q^\nu \Bigg\}
(\rlap/p_1 + m_1)~,\nnb \\
\es i {\lambda_{B_1} \lambda_{B_2}\over (p^2-m_2^2)
[(p+q)^2-m_1^2]} \Big\{ \rlap/p \rlap/\varepsilon \rlap/q (f_1+f_2) +
2 (\varepsilon\mcdot p) \rlap/p f_1  +(m_1-m_2) \rlap/p \rlap/\varepsilon \nnb \\
\ar 2 m_2 (\varepsilon\mcdot p) + (m_1 m_2 - p^2) \rlap/\varepsilon f_1
+ {f_2 \over m_1+m_2} \Big[ \rlap/p \rlap/\varepsilon \Big(
(p+q)^2-p^2 \Big) - 2 (\varepsilon\mcdot p) \rlap/p \rlap/q \nnb \\
\ar (p^2 + m_1 m_2) \rlap/\varepsilon \rlap/q + m_2 \Big( (p+q)^2-p^2 \Big)
\rlap/\varepsilon - 2 m_1 (\varepsilon\mcdot p) \rlap/q \Big] \Bigg\} \nnb \\
\es \Pi^{f_1+f_2} \rlap/p \rlap/\varepsilon \rlap/q + \Pi^{f_1} \rlap/p
(\varepsilon \cdot p) + \cdots~,
\eea
where we had set $p_1=p$ and $p_2=p+q$.

We see from Eq. (\ref{eoc12}) that the correlation function contains
numerous structures and none of the structures has any apparent advantage over
any other. Therefore any of these structures, in principle, can be used
in determining the baryon--meson coupling constants. Our numerical analysis
shows that the structures $\rlap/p \rlap/\varepsilon \rlap/q$ and $\rlap/p
(\varepsilon\mcdot p)$ exhibit better convergence, which is the reason why
we choose them in further analysis. From the coefficient functions $\Pi^{f_1+f_2}$ and
$\Pi^{f_1}$ one can extract the values of $f_1+f_2$ and $f_1$ respectively.

In order to obtain the expressions for
the correlation functions, and from which the coefficient functions, from the QCD side, baryon interpolating currents are
needed. In the present work we use the most general forms of the following
interpolating currents for baryons:
\bea
\label{eoc13}
\eta^{\Sigma^0} \es \sqrt{1\over 2} \epsilon^{abc} \Big[ \Big(u^{aT} C
s^b\Big)\gamma_5 d^c - \Big(s^{aT} C d^b\Big)\gamma_5 u^c +
\beta \Big( u^{aT} C \gamma_5 s^b\Big) d^c - \beta \Big( s^{aT} C \gamma_5
d^b\Big) u^c \Big]~, \nnb \\
\eta^{\Sigma^+} \es - \sqrt{1\over 2}\eta^{\Sigma^0}(d \rar u)~,\nnb \\
\eta^{\Sigma^-} \es -  \sqrt{1\over 2}\eta^{\Sigma^0}(u \rar d)~,\nnb \\
\eta^p \es  - \eta^{\Sigma^+}(s \rar d)~,\nnb \\
\eta^n \es  - \eta^{\Sigma^-}(s \rar u)~,\nnb \\
\eta^{\Xi^0} \es - \eta^n (d \rar s)~,\nnb \\
\eta^{\Xi^-} \es - \eta^p (u \rar s)~,\nnb \\
\eta^{\Lambda} \es - \sqrt{1\over 6} \epsilon^{abc} \Big[ 2 \Big(u^{aT} C
d^b\Big)\gamma_5 s^c + \Big(u^{aT} C s^b\Big)\gamma_5 d^c +
\Big( s^{aT} C \gamma_5 d^b\Big) u^c +2 \beta \Big( u^{aT} C \gamma_5   
d^b\Big) s^c \nnb \\
\ar \beta \Big( u^{aT} C \gamma_5 s^b\Big) d^c +
\Big( s^{aT} C \gamma_5 d^b\Big) u^c\Big]~, \nnb \\
\eea
where $C$ is the charge conjugation operator, $(a,b,c)$ are the color indices
and $\beta$ is an arbitrary parameter and $\beta=-1$ corresponds to
the Ioffe current. We see from Eq. (\ref{eoc13}) that all currents, except
the current of $\Lambda$, can be derived from the $\Sigma^0$ current by
making simple replacements. It is shown in \cite{Roc11} that
$\Lambda$ current can also be obtained from $\Sigma^0$ current with the help
of following relations:
\bea
\label{eoc14}
2 \eta^{\Sigma^0} (d \rar s) + \eta^{\Sigma^0} \es  \sqrt{3} \Lambda~, \nnb \\
2 \eta^{\Sigma^0} (u \rar s) - \eta^{\Sigma^0} \es -\sqrt{3} \Lambda~.
\eea
Before giving detailed calculations of the correlation functions, let us
firstly derive the relations among them. For this purpose we will follow the
approach presented in \cite{Roc07}, where relations between correlation
functions involving coupling constants of pseudoscalar mesons to octet
baryons are obtained. Of course, in the exact $SU(3)_f$ limit all coupling 
constants of vector mesons with octet baryons can be related to each other 
using symmetry arguments. The main advantage of our approach is that our 
approach allows us to take $SU(3)_f$ symmetry violating effects into account.

Below we will show that all correlation functions responsible for the
coupling constants of the vector mesons to octet baryons can be written in
terms of only three functions for each electric and magnetic form factors.
Note that the relations between the invariant functions are all
structure independent. Starting from the correlation function
that is responsible for the $\Sigma^0 \rar \Sigma^0 \rho^0$ transition, 
two of the three independent functions can be obtained. It allows us to
establish relations among this correlation function and the correlation
functions responsible for $\Sigma^+ \rar \Sigma^+ \rho^0$ and $\Sigma^- \rar
\Sigma^- \rho^0$ transitions, which can be written as:
\bea
\label{eoc15}
\Pi^{\Sigma^0 \rar \Sigma^0 \rho^0} = g_{\rho\bar{u}u} \Pi_1(u,d,s) +
g_{\rho\bar{d}d} \Pi_1^\prime(u,d,s) +
g_{\rho\bar{s}s} \Pi_2(u,d,s)~,
\eea
where we formally write down the quark content of the $\rho^0$ meson in the
form
\bea
\label{nolabel1}  
J_\mu = \sum_{u,d,s}  g_{\rho\bar{q}q} \bar{q} \gamma_\mu q~,\nnb
\eea
and for the $\rho^0$ meson $g_{\rho\bar{u}u} = - g_{\rho\bar{d}d}
= 1/\sqrt{2}$, $g_{\rho\bar{s}s} = 0$. The invariant functions $\Pi_1$,
$\Pi_1^\prime$ and $\Pi_2$ describe emission of the $\rho^0$ meson from
$u$, $d$ and $s$ quarks of $\Sigma^0$, respectively. We see from Eq.
(\ref{eoc13}) that the current of $\Sigma^0$ is symmetric under the
replacement $u \leftrightarrow d$, hence $\Pi_1^\prime(u,d,s) =
\Pi_1(d,u,s)$. For this reason, we have two independent functions
$\Pi_1(u,d,s)$ and $\Pi_2(u,d,s)$. In further discussion, we introduce the
following formal notation,
\bea
\label{eoc16}
\Pi_1(u,d,s) \es \lla \bar{u} u \vel \Sigma^0 \bar{\Sigma}^0 \ver 0\rra~, \nnb \\
\Pi_2(u,d,s)  \es \lla \bar{s} s \vel \Sigma^0 \bar{\Sigma}^0 \ver 0\rra~,
\eea
for convenience.  Replacing $d \rar u$ in $\Pi_1(d,u,s)$ and using
$\Sigma^0(d \rar u) = - \sqrt{2}\Sigma^+$, we obtain
\bea
\label{eoc17}
 4 \Pi_1(u,u,s) = 2 \lla \bar{u} u \vel \Sigma^+ \bar{\Sigma}^+ \ver 0\rra~.
\eea
Appearance of the factor 4 in Eq. (\ref{eoc17}) can be explained as follows.
Since there are two $u$ quarks, in $\Sigma^+$, there are two ways of
contracting them. Each one of the quark lines can emit the $\rho^0$ meson
yielding, in total, 4 ways for emitting the $\rho^0$
meson. Using the fact that in $\Sigma^+$ there is no $d$ quark, we obtain:
\bea
\label{eoc18}
\Pi^{\Sigma^+ \rar \Sigma^+ \rho^0} \es g_{\rho\bar{u}u} \lla \bar{u} u \vel
\Sigma^+ \bar{\Sigma}^+ \ver 0\rra + g_{\rho\bar{s}s} 
\lla \bar{s} s \vel \Sigma^+ \bar{\Sigma}^+ \ver 0\rra \nnb \\
\es \sqrt{2} \Pi_1(u,u,s)~.
\eea
Using similar arguments, for the $\Sigma^- \rar \Sigma^- \rho^0$ transition,
we get
\bea
\label{eoc19}
\Pi^{\Sigma^- \rar \Sigma^- \rho^0} \es g_{\rho\bar{d}d} \lla \bar{d} d \vel
\Sigma^- \bar{\Sigma}^- \ver 0\rra + g_{\rho\bar{s}s} 
\lla \bar{s} s \vel \Sigma^- \bar{\Sigma}^- \ver 0\rra \nnb \\
\es - \sqrt{2} \Pi_1^\prime(d,d,s) = - \sqrt{2} \Pi_1(d,d,s)~.
\eea
These equations establish relations between the couplings of $\rho^0$ meson
with $\Sigma^+$, $\Sigma^0$ and $\Sigma^-$ baryons. Note that in the isospin
symmetry limit, we obtain the well known relations
$\Pi^{\Sigma^+ \to \Sigma^+ \rho^0} = - \Pi^{\Sigma^- \to \Sigma^- \rho^0}$,
and $\Pi^{\Sigma^0 \to \Sigma^0 \rho^0} = 0$.   

Let us proceed now by calculating the couplings of $\rho^0$ meson with
proton and neutron. For this purpose we need the matrix elements
$\lla \bar{u} u \vel \bar{N}N \ver 0 \rra$ and $\lla \bar{d} d \vel \bar{N}
N\ver 0 \rra$. The matrix involving interpolating current of the
proton can be obtained from the current of $\Sigma^+$ by the replacement $s
\rar d$, i.e.,
\bea
\label{eoc20}
\lla \bar{u} u \vel p\bar{p} \ver 0\rra = \lla \bar{u} u \vel \Sigma^+
\bar{\Sigma}^+ \ver 0\rra (s\to d)= 2 \Pi_1(u,u,d)~.
\eea
In order to obtain $\lla \bar{d} d \vel p\bar{p} \ver 0\rra$,
$\Pi_2(u,d,s)$ is needed. In the first step, making the replacement $d \rar
u$, we get
\bea
\label{eoc21}
\Pi_2(u,u,s) = \lla \bar{s} s \vel \Sigma^+ \Sigma^- \ver 0\rra~,
\eea
where the factor 2 in the normalization of the current is canceled by the
two possible ways of contracting the $u$ quarks.
Making the replacement $s \rar d$ in the second step, we obtain
\bea
\label{eoc22}
\Pi_2(u,u,d) = \lla \bar{d} d \vel p\bar{p} \ver 0\rra~.
\eea
It follows from Eqs. (\ref{eoc20})--(\ref{eoc22}) that
\bea
\label{eoc23}
\Pi^{p \rar p \rho^0} \es g_{\rho\bar{u}u} \lla \bar{u} u \vel
p \bar{p} \ver 0\rra + g_{\rho\bar{d}d}
\lla \bar{d} d \vel p \bar{p} \ver 0\rra \nnb \\
\es \sqrt{2} \Pi_1(u,u,d) - {1\over\sqrt{2}} \Pi_2(u,u,d)~.
\eea
Similarly, we can easily obtain the following results involving the coupling
constants of $\rho^0$ meson to the neutron and $\Xi$ baryons,
\bea
\label{eoc24}
\Pi^{n \rar n \rho^0} \es {1\over\sqrt{2}} \Pi_2(d,d,u) - \sqrt{2}
\Pi_1(d,d,u)~, \nnb \\ 
\Pi^{\Xi^0 \rar \Xi^0 \rho^0} \es {1\over\sqrt{2}} \Pi_2(s,s,u)~, \nnb \\
\Pi^{\Xi^- \rar \Xi^- \rho^0} \es -{1\over\sqrt{2}} \Pi_2(s,s,d)~.
\eea
These relations, together with the relations given in 
Eqs. (\ref{eoc18}), (\ref{eoc19}) and (\ref{eoc23}), describe the couplings
of $\rho^0$ meson with baryons in terms of two invariant functions
$\Pi_1(u,d,s)$ and $\Pi_2(u,d,s)$.

Now let us derive similar relations for the charged $\rho$ meson. Consider
the matrix element $\lla \bar{d} d \vel \Sigma^0 \bar{\Sigma}^0 \ver 0\rra$
in which $d$ quarks from the $\Sigma^0$ and $\bar{\Sigma}^0$ form the final
$\bar{d}d$ and the other $u$ and $s$ quarks are the spectators. In the
matrix element $\lla \bar{u} d \vel \Sigma^+ \bar{\Sigma}^0 \ver 0\rra$, $d$
quark from $\Sigma^0$ and $u$ quark from $\Sigma^+$ form the state
$\bar{u}d$ and the other $u$ and $s$ quarks, similar to the previous case,
are the spectators. Therefore, one can expect that these matrix elements
should be proportional, and indeed, calculations confirm that, i.e.,
\bea
\label{eoc25}
\Pi^{\Sigma^0 \rar \Sigma^+ \rho^-} \es \lla \bar{u} d \vel
\Sigma^+ \bar{\Sigma}^0 \ver 0\rra
= - \sqrt{2} \lla \bar{d} d \vel \Sigma^0 \bar{\Sigma}^0 \ver 0\rra
= - \sqrt{2} \Pi_1^\prime(u,d,s) \nnb \\
\es - \sqrt{2} \Pi_1(d,u,s)~.
\eea
Making the exchange $u \leftrightarrow d$ in Eq. (\ref{eoc25}), we get
\bea
\label{eoc26}
\Pi^{\Sigma^0 \rar \Sigma^- \rho^+} \es \lla \bar{d} u \vel
\Sigma^- \bar{\Sigma}^0 \ver 0\rra
= \sqrt{2} \lla \bar{u} u \vel \Sigma^0 \bar{\Sigma}^0 \ver 0\rra
= \sqrt{2} \Pi_1(u,d,s)~.
\eea
Performing similar calculations for the $\Xi$ baryons, we obtain:
\bea
\label{eoc27}                 
\Pi^{\Xi^0 \rar \Xi^- \rho^+} \es \lla \bar{d} u \vel
\Xi^- \bar{\Xi}^0 \ver 0\rra
= - \sqrt{2} \lla \bar{u} u \vel \Xi^0 \bar{\Xi}^0 \ver 0\rra
=  \Pi_2(s,s,u)~, \nnb \\
\Pi^{\Xi^- \rar \Xi^0 \rho^-} \es \lla \bar{u} d \vel
\Xi^0 \bar{\Xi}^- \ver 0\rra         
= - \sqrt{2} \lla \bar{d} d \vel \Xi^0 \bar{\Xi}^0 \ver 0\rra        
=  \Pi_2(s,s,d)~.
\eea

The correlation functions involving the $\rho$ and $K^\ast$ mesons can be
written as:
\bea
\label{eoc28}
\Pi^{\Sigma^- \rar \Sigma^0 \rho^-} \es \sqrt{2} \Pi_1(u,d,s)~, \nnb \\
\Pi^{\Sigma^+ \rar \Sigma^0 \rho^+} \es \sqrt{2} \Pi_1^\prime(u,d,s) =
- \sqrt{2} \Pi_1(d,u,s) ~, \nnb \\
\Pi^{\Sigma^- \rar n K^{\ast -}} \es - \Pi_2(d,d,s)~, \nnb \\
\Pi^{p \rar \Sigma^+ K^{\ast 0}} \es - \Pi_2(u,u,d)~, \nnb \\
\Pi^{\Sigma^+ \rar p \bar{K}^{\ast 0}} \es - \Pi_2(u,u,s)~, \nnb \\
\Pi^{n \rar \Sigma^- K^{\ast +}} \es - \Pi_2(d,d,s)~.
\eea

We make use of Eqs. (\ref{eoc14}) in order to calculate 
the correlation functions involving the $\Lambda$ baryon, in terms of 
the invariant functions, after which we get:
\bea
\label{eoc29}
\Pi^{\Lambda \rar \Lambda \rho^0} \es {\sqrt{2}\over 3} 
\Big[ \Pi_1(u,s,d) -  \Pi_1(d,s,u) + \Pi_2(s,d,u) -\Pi_2(s,u,d) \nnb \\
\ek {1\over 2} \Pi_1(u,d,s)
+ {1\over 2} \Pi_1(d,u,s) \Big] ~, \nnb \\
\Pi^{\Lambda \rar \Sigma^0 \rho^0} + \Pi^{\Sigma^0 \rar \Lambda \rho^0} \es 
{2\over \sqrt{6}} \Big[ \Pi_1(u,s,d) +  \Pi_1(d,s,u) - \Pi_2(s,d,u) 
-\Pi_2(s,u,d) \Big] ~, \nnb \\
\Pi^{\Xi^- \rar \Sigma^0 K^{\ast -}} + \sqrt{3}\Pi^{\Xi^- \rar \Lambda
K^{\ast -}} \es - 2 \sqrt{2} \Pi_1(u,s,d) ~, \nnb \\
\Pi^{n \rar \Sigma^0 K^{\ast 0}} - \sqrt{3}\Pi^{n \rar \Lambda
K^{\ast 0}} \es  2 \sqrt{2} \Pi_1(s,d,u) ~, \nnb \\
\Pi^{p \rar \Sigma^0 K^{\ast +}} + \sqrt{3}\Pi^{p \rar \Lambda
K^{\ast +}} \es  - 2 \sqrt{2} \Pi_1(s,u,d) ~, \nnb \\
- \Pi^{\Xi^0 \rar \Sigma^0 K^{\ast 0}} + \sqrt{3}\Pi^{\Xi^0 \rar \Lambda
K^{\ast 0}} \es  2 \sqrt{2} \Pi_1(d,s,u) ~, \nnb \\
\Pi^{\Sigma^0 \rar p K^{\ast -}} + \sqrt{3}\Pi^{\Lambda \rar p
K^{\ast -}} \es  - 2 \sqrt{2} \Pi_1(s,u,d) ~, \nnb \\
\Pi^{\Sigma^0 \rar n K^{\ast 0}} - \sqrt{3}\Pi^{\Lambda \rar n      
K^{\ast 0}} \es  2 \sqrt{2} \Pi_1(s,d,u) ~, \nnb \\
\Pi^{\Sigma^0 \rar \Xi^0 \bar{K}^{\ast 0}} - \sqrt{3}\Pi^{\Lambda \rar \Xi^0      
\bar{K}^{\ast 0}} \es  - 2 \sqrt{2} \Pi_1(d,s,u) ~, \nnb \\
\Pi^{\Sigma^0 \rar \Xi^- K^{\ast +}} + \sqrt{3}\Pi^{\Lambda \rar \Xi^-      
K^{\ast +}} \es  2 \sqrt{2} \Pi_1(u,s,d) ~.
\eea

As can easily be seen in Eq. (\ref{eoc29}), correlation functions involving
a single $\Lambda$ baryon always come together with correlation functions
involving a $\Sigma^0$ baryon, and therefore it is impossible to separate 
them using only $\Pi_1$ and $\Pi_2$. To be able to separate the correlation functions 
involving the $\Lambda$ and $\Sigma^0$ baryons, we need to introduce one more 
independent function
\bea
\label{eoc30}
\Pi_3(u,d,s) \es - \Pi^{\Sigma^0 \rar \Xi^- K^{\ast +}} = - \lla u \bar{s} \vel
\Xi^- \bar{\Sigma}^0 \ver 0\rra~.
\eea
Note that in \cite{Roc07}, a fourth function is defined as
\bea
\label{eoc31}
\Pi_4(u,d,s) \es - \Pi^{\Xi^- \rar \Sigma^0 K^{\ast -}} = - \lla s \bar{u} \vel
\Sigma^0 \bar{\Xi}^- \ver 0\rra~.
\eea
In the present work a new relation missed in \cite{Roc07} is obtained, namely,
$\Pi_4(u,d,s) = \Pi_3(u,s,d)$, and hence
such a fourth function is not necessary. Choice of  $\Pi_3(u,d,s)$ is not unique.

Using this new invariant function and performing simple calculations, we
obtain:
\bea
\label{eoc32}
\Pi^{\Xi^0 \rar \Sigma^+ K^{\ast -}} \es \Pi^{n \rar p \rho^-} (d
\rar s) = - \sqrt{2} \Pi_3(s,s,u) ~, \nnb \\
\Pi^{\Xi^- \rar \Sigma^- K^{\ast 0}} \es \Pi^{\Xi^0 \rar \Sigma^+  K^{\ast
-}} (u \rar d) = - \sqrt{2} \Pi_3(s,s,d) ~, \nnb \\
\Pi^{\Sigma^+ \rar \Xi^0 K^{\ast +}} \es \sqrt{2} \Pi^{\Sigma^0 \rar
\Xi^-  K^{\ast +}} (d \rar u) = - \sqrt{2} \Pi_3(u,u,s) ~, \nnb \\
\Pi^{p \rar n \rho^+} \es \Pi^{\Sigma^+ \rar
\Xi^0  K^{\ast +}} (s \rar d) = - \sqrt{2} \Pi_3(u,u,d) ~, \nnb \\
\Pi^{n \rar p \rho^-} \es \Pi^{p \rar n  \rho^+} (u \lrar d) = 
- \sqrt{2} \Pi_3(d,d,u) ~, \nnb \\
\Pi^{\Sigma^0 \rar p K^{\ast -}} - 
\sqrt{3} \Pi^{\Lambda \rar p K^{\ast -}}
\es 2 \Pi^{\Sigma^0 \rar \Xi^- K^{\ast +}}(s \lrar u) =
- 2 \Pi_3(s,d,u) ~, \nnb \\
\Pi^{\Sigma^0 \rar n K^{\ast 0}} + 
\sqrt{3} \Pi^{\Lambda \rar n K^{\ast 0}}
\es -2 \Pi^{\Sigma^0 \rar \Xi^- K^{\ast +}}(u \lrar s)(d \lrar u) =
2 \Pi_3(s,u,d) ~, \nnb \\
\Pi^{\Sigma^0 \rar \Sigma^- \rho^+} + 
\sqrt{3} \Pi^{\Lambda \rar \Sigma^- \rho^+}
\es 2 \Pi^{\Sigma^0 \rar \Xi^- K^{\ast +}}(s \lrar d) =
-2 \Pi_3(u,s,d) ~, \nnb \\
\Pi^{\Lambda \rar \Sigma^+ \rho^-} \es \Pi^{\Lambda \rar \Sigma^- \rho^+}
(u \rar d)~, \nnb \\
\Pi^{\Sigma^- \rar \Xi^- \bar{K}^{\ast 0}} \es \Pi^{\Sigma^+ \rar \Xi^0
K^{\ast +}} (u \rar d) = - \sqrt{2} \Pi_3(d,d,s) ~, \nnb \\
\Pi^{\Sigma^0 \rar \Xi^0 \bar{K}^{\ast 0}} \es - \Pi^{\Sigma^0 \rar \Xi^-
K^{\ast +}} (d \lrar u) = \Pi_3(d,u,s) ~, \nnb \\
\Pi^{\Xi^0 \rar \Sigma^0 K^{\ast 0}} \es - \Pi^{\Xi^- \rar \Sigma^0
K^{\ast -}} (u \rar d) = \Pi_3(d,s,u)~, \nnb \\
\Pi^{p \rar \Sigma^0 K^{\ast +}} - \sqrt{3} \Pi^{p \rar \Lambda K^{\ast +}}
\es 2 \Pi^{\Xi^- \rar \Sigma^0 K^{\ast -}} (u \lrar s) = 
- 2 \Pi_3(s,u,d) ~,
\nnb \\
\Pi^{n \rar \Sigma^0 \bar{K}^{\ast 0}} + \sqrt{3} \Pi^{n \rar \Lambda
\bar{K}^{\ast 0}}
\es - 2 \Pi^{\Xi^- \rar \Sigma^0 K^{\ast -}} (u \lrar s)(d \lrar u) = 
2 \Pi_3(s,d,u) ~, \nnb \\
\Pi^{\Sigma^- \rar \Sigma^0 \rho^-} + \sqrt{3} \Pi^{\Sigma^- \rar \Lambda
\rho^-} \es  2 \Pi^{\Xi^- \rar \Sigma^0 K^{\ast -}} (d \rar s) = 
- 2 \Pi_3(u,d,s) ~, \nnb \\
\Pi^{\Sigma^+ \rar \Lambda \rho^+} \es \Pi^{\Sigma^- \rar \Lambda \rho^-}
(u \lrar d)  ~, \nnb \\
\Pi^{\Sigma^0 \rar \Sigma^0 \omega} \es {1\over \sqrt{2}} \Big[
\Pi_1(u,d,s) + \Pi_1(d,u,s) \Big] ~, \nnb \\
\Pi^{\Sigma^+ \rar \Sigma^+ \omega} \es \sqrt{2} \Pi_1(u,u,s) ~, \nnb \\
\Pi^{\Sigma^- \rar \Sigma^- \omega} \es \sqrt{2} \Pi_1(d,d,s) ~, \nnb \\
\Pi^{p \rar p \omega} \es \sqrt{2} \Pi_1(u,u,d) + {1\over \sqrt{2}}
\Pi_2(u,u,d)~, \nnb \\
\Pi^{n \rar n \omega} \es \sqrt{2} \Pi_1(d,d,u) + {1\over \sqrt{2}}
\Pi_2(d,d,u)~, \nnb \\
\Pi^{\Xi^0 \rar \Xi^0 \omega} \es {1\over \sqrt{2}} \Pi_2(s,s,u) ~, \nnb \\
\Pi^{\Xi^- \rar \Xi^- \omega} \es {1\over \sqrt{2}} \Pi_2(s,s,d) ~, \nnb \\
\Pi^{\Sigma^0 \rar \Lambda \omega} + \Pi^{\Lambda \rar \Sigma^0 \omega} \es
{2\over \sqrt{6}} \Big[\Pi_1(u,s,d) - \Pi_1(d,s,u) - \Pi_2(s,d,u) +
\Pi_2(s,u,d)\Big] ~, \nnb \\
\Pi^{\Lambda \rar \Lambda \omega} \es
{\sqrt{2}\over 3} \Bigg[\Pi_1(u,s,d) + \Pi_1(d,s,u) + \Pi_2(s,d,u) +
\Pi_2(s,u,d) \nnb \\
\ek {1\over 2} \Pi_1(u,d,s) - {1\over 2} \Pi_1(d,u,s) \Bigg] ~, \nnb \\
\Pi^{\Sigma^0 \rar \Sigma^0 \phi} \es \Pi_2(u,d,s) ~, \nnb \\
\Pi^{\Sigma^+ \rar \Sigma^+ \phi} \es \Pi_2(u,u,s) ~, \nnb \\
\Pi^{\Sigma^- \rar \Sigma^- \phi} \es \Pi_2(d,d,s) ~, \nnb \\
\Pi^{p \rar p \phi} \es \Pi^{n \rar n \phi} = 0~, \nnb \\
\Pi^{\Xi^0 \rar \Xi^0 \phi} \es 2 \Pi_1(s,s,u) ~, \nnb \\
\Pi^{\Xi^- \rar \Xi^- \phi} \es 2 \Pi_1(s,s,d) ~, \nnb \\
\Pi^{\Sigma^0 \rar \Lambda \phi} + \Pi^{\Lambda \rar \Sigma^0 \phi} \es
{2\over \sqrt{3}} \Big[ - \Pi_1(s,d,u) + \Pi_1(s,u,d) \Big] ~, \nnb \\
\Pi^{\Lambda \rar \Lambda \phi} \es {2\over 3} \Bigg[\Pi_1(s,d,u) +
\Pi_1(s,u,d) - {1\over 2} \Pi_2(u,d,s)\Bigg]~.
\eea  

These relations allow us to express, all possible strong coupling
constants of the octet vector mesons with the octet baryons in terms of three independent invariant functions without using
the flavor symmetry.

Before starting to calculate these invariant functions from the QCD side we
would like to make the following remark. The invariant function
$\Pi_3(u,d,s)$ can be split into symmetric and antisymmetric parts with
respect to the exchange of $d$ and $s$ quarks as:
\bea
\Pi_3(u,d,s) = \Pi_3^{sym}(u,d,s) + \Pi_3^{asym}(u,d,s)~. \nnb 
\eea
The symmetric part, $\Pi_3^{sym}$, can be expressed in terms of
$\Pi_1$ and $\Pi_2$ as 
\bea
\Pi_3^{sym}(u,d,s) = {1\over \sqrt{2}} \Big[ \Pi_1(u,d,s) + \Pi_1(u,s,d) -
\Pi_2(s,d,u) \Big]~.\nnb
\eea
The explicit form of $\Pi_3^{asym}(u,d,s)$, which vanishes in the $SU(3)_f$
limit, is given in the appendix along
with the explicit forms of $\Pi_1$ and $\Pi_2$. 
Hence, in $SU(3)_f$ limit
only two invariant functions $\Pi_1$ and $\Pi_2$ are relevant and they
correspond to ${\cal F}$ and ${\cal D}-{\cal F}$ couplings. One more 
additional function $\Pi_3^{asym}$ is needed in order to take $SU(3)_f$ 
violation into consideration.
  
We can now proceed to calculate these three invariant functions. As has
already been mentioned, for this aim, the correlation functions responsible 
for the transitions $\Sigma^0 \rar \Sigma^0 \rho^0$, $\Sigma^0 \rar 
\Sigma^0 \phi$ and  $\Sigma^0 \rar \Xi^- K^{\ast +}$ are enough
(see Eq. (\ref{eoc08})).

In deep Euclidean region, $-p_1^2 \rar \infty$, $-p_2^2 \rar \infty$, the
correlation function can be evaluated using OPE. In order to obtain the
expressions of the correlation functions from QCD side, the propagator of
the light quarks and the matrix elements of the nonlocal operators
$\bar{q}(x_1) \Gamma q^\prime (x_2)$ and $\bar{q}(x_1) G_{\mu\nu} q^\prime
(x_2)$ between the vacuum and the vector meson states are needed, where
$\Gamma$ represents the Dirac matrices relevant to the case under
consideration, and $G_{\mu\nu}$ is the gluon field strength tensor. 

Up to twist--4 accuracy,
matrix elements $\lla V(q) \vel \bar{q}(x) \Gamma q(0) \ver 0 \rra$ and 
$\lla V(q) \vel \bar{q}(x) G_{\mu\nu} q(0) \ver 0 \rra$ are given in
\cite{Roc12,Roc13} as follows:
\bea
\label{eoc33}
\lla V(q,\lambda) \vel \bar{q}_1(x) \gamma_\mu q_2(0) \ver 0 \rra \es
f_V m_V \Bigg\{ {\varepsilon^\lambda \mcdot x \over q\mcdot x} q_\mu 
\int_0^1 du e^{i \bar{u} q\mcdot x} \Bigg[ \phi_\parallel (u) + {m_V^2 x^2 \over 16}
A_\parallel (u) \Bigg] \nnb \\
\ar \Bigg( \varepsilon_\mu^\lambda - q_\mu {\varepsilon^\lambda \mcdot x \over q\mcdot x} \Bigg) \int_0^1
du e^{i \bar{u} q\mcdot x} g_\perp^v (u) \nnb \\
\ek {1\over 2} x_\mu { \varepsilon^\lambda \mcdot x
\over (q\mcdot x)^2} m_V^2 \int_0^1 du e^{i \bar{u} q\mcdot x} \Big[ g_3 (u) +
\phi_\parallel (u) - 2 g_\perp^v (u) \Big] \Bigg\}~, \nnb \\ \nnb \\
\lla V(q,\lambda) \vel \bar{q}_1(x) \gamma_\mu \gamma_5 q_2(0) \ver 0 \rra
\es - {1 \over 4} \epsilon_\mu^{\nu\alpha\beta} \varepsilon^\lambda q_\alpha
x_\beta f_V m_V \int_0^1 du e^{i \bar{u} q\mcdot x} g_\perp^a (u)~, 
\nnb \\ \nnb \\
\lla V(q,\lambda) \vel \bar{q}_1(x) \sigma_{\mu\nu} q_2(0) \ver 0 \rra \es  
- i f_V^T \Bigg\{ (\varepsilon_\mu^\lambda q_\nu - \varepsilon_\nu^\lambda
q_\mu ) \int_0^1 du e^{i \bar{u} q\mcdot x}\Bigg[\phi_\perp (u) + {m_V^2
x^2 \over 16} A_\perp (u) \Bigg] \nnb \\
\ar { \varepsilon^\lambda \mcdot x \over (q\mcdot x)^2} (q_\mu x_\nu - q_\nu
x_\mu) \int_0^1 du e^{i \bar{u} q\mcdot x} \Bigg[h_\parallel^t - {1\over 2}
\phi_\perp - {1\over 2} h_3 (u) \Bigg] \nnb \\
\ar {1\over 2} (\varepsilon_\mu^\lambda x_\nu - \varepsilon_\nu^\lambda
x_\mu) {m_V^2 \over q \mcdot x} \int_0^1 du e^{i \bar{u} q\mcdot x}
\Big[h_3(u) - \phi_\perp (u) \Big] \Bigg\}~, \nnb \\ \nnb \\
\lla V(q,\lambda) \vel \bar{q}_1(x) \sigma_{\alpha\beta} g G_{\mu\nu}(u x) q_2(0) \ver 0 \rra \es  
f_V^T m_V^2 { \varepsilon^\lambda \mcdot x \over 2 q\mcdot x} \Big[q_\alpha q_\mu
g_{\beta\nu}^\perp - q_\beta q_\mu g_{\alpha\nu}^\perp - q_\alpha q_\nu g_{\beta\mu}^\perp
+ q_\beta q_\nu g_{\alpha\mu}^\perp \Big] \nnb \\
\cp \int {\cal D}\alpha_i e^{i(\alpha_{\bar{q}}
+ u \alpha_g) q\mcdot x} {\cal T}(\alpha_i) \nnb \\ 
\ar f_V^T m_V^2 \Big[q_\alpha \varepsilon_\mu^\lambda g_{\beta\nu}^\perp - q_\beta
\varepsilon_\mu^\lambda g_{\alpha\nu}^\perp - q_\alpha
\varepsilon_\nu^\lambda g_{\beta\mu}^\perp
+ q_\beta \varepsilon_\nu^\lambda g_{\alpha\mu}^\perp \Big] \nnb \\
\cp \int {\cal D}\alpha_i e^{i(\alpha_{\bar{q}}
+ u \alpha_g) q\mcdot x} {\cal T}_1^{(4)}(\alpha_i) \nnb \\
\ar f_V^T m_V^2 \Big[q_\mu \varepsilon_\alpha^\lambda g_{\beta\nu}^\perp -
q_\mu \varepsilon_\beta^\lambda g_{\alpha\nu}^\perp - q_\nu
\varepsilon_\alpha^\lambda g_{\beta\mu}^\perp
+ q_\nu \varepsilon_\beta^\lambda g_{\alpha\mu}^\perp \Big] \nnb \\
\cp \int {\cal D}\alpha_i e^{i(\alpha_{\bar{q}}
+ u \alpha_g) q\mcdot x} {\cal T}_2^{(4)}(\alpha_i) \nnb \\
\ar {f_V^T m_V^2 \over q \mcdot x} \Big[q_\alpha q_\mu \varepsilon_\beta^\lambda
x_\nu - q_\beta q_\mu \varepsilon_\alpha^\lambda x_\nu -
q_\alpha q_\nu \varepsilon_\beta^\lambda x_\mu +
q_\beta q_\nu \varepsilon_\alpha^\lambda x_\mu \nnb \\
\cp \int {\cal D}\alpha_i e^{i(\alpha_{\bar{q}}
+ u \alpha_g) q\mcdot x} {\cal T}_3^{(4)}(\alpha_i) \nnb \\
\ar {f_V^T m_V^2 \over q \mcdot x} \Big[q_\alpha q_\mu \varepsilon_\nu^\lambda
x_\beta - q_\beta q_\mu \varepsilon_\nu^\lambda x_\alpha -
q_\alpha q_\nu \varepsilon_\mu^\lambda x_\beta +
q_\beta q_\nu \varepsilon_\mu^\lambda x_\alpha \nnb \\
\cp \int {\cal D}\alpha_i e^{i(\alpha_{\bar{q}}
+ u \alpha_g) q\mcdot x} {\cal T}_4^{(4)}(\alpha_i)~, \nnb \\ \nnb \\
\lla V(q,\lambda) \vel \bar{q}_1(x) g_s G_{\mu\nu} (ux) q_2(0) \ver 0 \rra \es
-i f_V^T m_V (\varepsilon_\mu^\lambda q_\nu - \varepsilon_\nu^\lambda q_\mu)
\int {\cal D}\alpha_i e^{i(\alpha_{\bar{q}}
+ u \alpha_g) q\mcdot x} {\cal S}(\alpha_i)~, \nnb \\ \nnb \\
\lla V(q,\lambda) \vel \bar{q}_1(x) g_s \widetilde{G}_{\mu\nu} (ux) 
\gamma_5 q_2(0) \ver 0 \rra \es
-i f_V^T m_V (\varepsilon_\mu^\lambda q_\nu - \varepsilon_\nu^\lambda q_\mu)
\int {\cal D}\alpha_i e^{i(\alpha_{\bar{q}}
+ u \alpha_g) q\mcdot x} \widetilde{{\cal S}}(\alpha_i)~, \nnb \\ \nnb \\
\lla V(q,\lambda) \vel \bar{q}_1(x) g_s \widetilde{G}_{\mu\nu} (ux) 
\gamma_\alpha \gamma_5 q_2(0) \ver 0 \rra \es
f_V m_V q_\alpha (\varepsilon_\mu^\lambda q_\nu - \varepsilon_\nu^\lambda q_\mu)
\int {\cal D}\alpha_i e^{i(\alpha_{\bar{q}}
+ u \alpha_g) q\mcdot x} {\cal A}(\alpha_i)~, \nnb \\ \nnb \\
\lla V(q,\lambda) \vel \bar{q}_1(x) g_s G_{\mu\nu} (ux) i \gamma_\alpha 
q_2(0) \ver 0 \rra \es
f_V m_V q_\alpha (\varepsilon_\mu^\lambda q_\nu - \varepsilon_\nu^\lambda q_\mu)
\int {\cal D}\alpha_i e^{i(\alpha_{\bar{q}}
+ u \alpha_g) q\mcdot x} {\cal V}(\alpha_i)~,
\eea
where $\widetilde{G}_{\mu\nu} = (1/2) \epsilon_{\mu\nu\alpha\beta}
G^{\alpha\beta}$ is the dual gluon field strength tensor, and $\int {\cal D}
\alpha_i = \int d\alpha_q d\alpha_{\bar{q}} d\alpha_g \delta (1 - \alpha_q -
\alpha_{\bar{q}} - \alpha_g)$. In further analysis, we use the following
expression for the light quark propagator
\bea
\label{eoc34}
S_q(x) \es {i \rlap/x \over 2 \pi^2 x^4} - {m_q \over 4 \pi^2 x^2} -
{\lla\bar{q}q\rra \over 12} \Bigg(1 - {i m_q \over 4} \rlap/x \Bigg) -
{x^2 \over 192} m_0^2 \lla\bar{q}q\rra  \Bigg(1 - {i m_q \over 6} 
\rlap/x \Bigg) \nnb \\
\ek i g_s \int_0^1 du \Bigg\{ {\rlap/x \over 16 \pi^2 x^2} G_{\mu\nu} (ux)
\sigma^{\mu\nu} - u x^\mu G_{\mu\nu} (ux) \gamma^\nu {i \over 4 \pi^2 x^2}
\nnb \\
\ek {i m_q \over 32 \pi^2} G_{\mu\nu} (ux) \sigma^{\mu\nu} \Bigg[ \ln \Bigg(
{- x^2 \Lambda^2 \over 4} \Bigg) + 2 \gamma_E \Bigg] \Bigg\}~,
\eea 
where $\gamma_E$ is the Euler constant, $\Lambda$ is a scale parameter, and
we will choose it as a factorization scale, i.e., $\Lambda = 0.5 \div 1.0~GeV$ 
(for more detail, see \cite{Roc14}). Note that, in the calculations, $SU(3)_f$
symmetry violation effects are included in the nonzero strange quark mass and
different strange quark condensate. These effects are also taken into
account in calculation of the distribution amplitudes \cite{Roc12,Roc13}.

The expressions of the full propagator of the light quark and the definition
of the distribution amplitudes allow us to calculate the theoretical part of
the correlation functions. Equating both representations of correlation
function and separating coefficients of Lorentz structures 
$\rlap/p \rlap/\varepsilon \rlap/q$ and $\rlap/p (\varepsilon \mcdot p)$ and
applying Borel transformation to both side of the correlation functions
on the variables $p^2$ and $(p+q)^2$ in order to suppress the
contributions of the higher states and continuum (see
\cite{Roc17}), we get the sum rules for the corresponding vector
meson baryon couplings. The contributions of higher states and the continuum are subtracted using quark-hadron duality.
After lengthy calculations, for each Lorentz
structure the expressions for the three invariant functions 
$\Pi_i^{(\alpha)},~i=1,2,3$ are obtained and their expressions are
presented in Appendix--A.
Here superscript $\alpha$ refers to the invariant functions
$\Pi_i^{(\alpha)}$ relevant to the coupling constants $f_1$ and $f_1+f_2$,
respectively.

For a given transition $B_1 \rightarrow B_2 V$, once the Borel transformed and continuum subtracted 
coefficient functions $\Pi^{f_1}$ and $\Pi^{f_1+f_2}$ are obtained,
the sum rules for the electric and magnetic type couplings are obtained as
\begin{eqnarray}
f_1 = \frac{1}{\lambda_{B_1} \lambda_{B_2}}e^{-{m_1^2 \over M_1^2} - {m_2^2 \over M_2^2} - {m_V^2 \over{M_1^2 + M_2^2}}} \Pi^{f_1}
\nonumber \\
f_1+f_2 = \frac{1}{\lambda_{B_1} \lambda_{B_2}}e^{-{m_1^2 \over M_1^2} - {m_2^2 \over M_2^2} - {m_V^2 \over{M_1^2 + M_2^2}}} \Pi^{f_1+f_2}
\nonumber 
\end{eqnarray}

In determining the vector meson--octet baryon strong coupling constants, the
residues of baryons are needed. The residues of baryons are obtained from
the analysis of two--point correlation function are given in
\cite{Roc15,Roc16,Roc17}. The currents of the other baryons can be obtained from
$\Sigma^0$ current by making appropriate substitutions of quarks. For this
reason, for determination of the residues, we give the sum rule only for
$\Sigma^0$
\bea
\label{eoc35}
\lambda_{\Sigma^0}^2 e^{-m_{\Sigma^0}^2/M^2} \es
{M^6\over 1024 \pi^2} (5 + 2 \beta + 5 \beta^2) - {m_0^2\over 96 M^2} (-1+\beta)^2 \lla
\bar{u} u \rra \lla \bar{d} d \rra \nnb \\
\ek {m_0^2\over 16 M^2} (-1+\beta^2) \lla \bar{s} s \rra \Big(\lla \bar{u} u \rra +
\lla \bar{d} d \rra\Big) \nnb \\
\ar {3 m_0^2\over 128} (-1+\beta^2) \Big[ m_s \Big(\lla \bar{u} u \rra +   
\lla \bar{d} d \rra\Big) + (m_u+m_d) \lla \bar{s} s \rra \Big] \nnb \\
\ek  {1\over 64 \pi^2} (-1+\beta)^2 M^2 \Big( m_d \lla \bar{u} u \rra + m_s \lla \bar{d} d
\rra \Big) \nnb \\
\ek {3 M^2\over 64 \pi^2} (-1+\beta^2) \Big[ m_s \Big(\lla \bar{u} u \rra +   
\lla \bar{d} d \rra\Big) + (m_u+m_d) \lla \bar{s} s \rra \Big] \nnb \\
\ar {1\over 128 \pi^2}  (5 + 2 \beta + 5 \beta^2) \Big( m_u \lla \bar{u} u \rra +
m_d \lla \bar{d} d \rra + m_s \lla \bar{s} s \rra\Big) \nnb \\
\ar {1\over 24} \Big[ 3 (-1+\beta^2) \lla \bar{s} s \rra \Big(\lla \bar{u} u
\rra + \lla \bar{d} d \rra \Big) + (-1+\beta^2) \lla \bar{u} u \rra \lla \bar{d}
d \rra \Big] \nnb \\
\ar {m_0^2\over 256 \pi^2} (-1+\beta)^2 \Big(m_u \lla \bar{d} d \rra +   
m_d \lla \bar{u} u \rra\Big) \nnb \\
\ar {m_0^2\over 26 \pi^2} (-1+\beta^2) \Big[ 13 m_s \Big(\lla \bar{u} u \rra +   
\lla \bar{d} d \rra\Big) + 11 (m_u+m_d) \lla \bar{s} s \rra \Big] \nnb \\
\ek {m_0^2\over 192 \pi^2}(1+\beta+\beta^2) \Big( m_u \lla \bar{u} u \rra +
m_d \lla \bar{d} d \rra - 2 m_s \lla \bar{s} s \rra\Big)~.  
\eea   
It follows from Eq. (\ref{eoc35}) that, sum rules cannot predict the sign
of the residue. We have chosen the sign convention such that in the
$SU(3)_f$ symmetry, the signs correctly reproduce the ${\cal F}$ and ${\cal
D}$ couplings (see \cite{Roc07}).

\section{Numerical analysis and discussion}

This section is devoted to the numerical analysis of the sum rules for the
vector meson--octet baryon coupling constants. The main input parameters of
the light cone sum rules in our case are the vector meson distribution
amplitudes (DAs). The DAs of the vector mesons are given in    
\cite{Roc12,Roc13,Roc14}. The values of the leptonic constants $f_V$ and $f_V^T$,
and of the twist--2 and twist--3 parameters $a_i^\parallel$,
$a_i^\perp$,$\zeta_{3V}^\parallel$, $\tilde{\lambda}_{3V}^\parallel$,
$\tilde{\omega}_{3V}^\parallel$, $\kappa_{3V}^\parallel$,
$\omega_{3V}^\parallel$, $\lambda_{3V}^\parallel$, $\kappa_{3V}^\perp$,
$\omega_{3V}^\perp$, $\lambda_{3V}^\perp$, as well as twist--4 parameters 
$\zeta_4^\parallel$, $\tilde{\omega}_4^\parallel$, $\zeta_4^\perp$,
 $\tilde{\zeta}_4^\perp$, $\kappa_{4V}^\parallel$, $\kappa_{4V}^\perp$ are
given in Table (1) and Table (2), respectively, in \cite{Roc12}.
The value of the other input parameters which are needed in the sum rule are
$\lla \bar{q} q \rra = - (0.243~GeV)^3$, $m_0^2 = 0.8$ \cite{Roc15},
$\GG = 0.47~GeV^4$ \cite{Roc03}.  

In the problem under consideration, the masses of initial and final baryons
are, more or less, equal to each other. For this reason we choose
$M_1^2=M_2^2=2 M^2$, and consequently we set $u_0=1/2$. Hence, in further
numerical analysis, the values of the DAs only at $u_0=1/2$ are needed.  
  
It follows from the explicit expressions of the sum rules for the vector 
meson--octet baryon coupling constants that, in addition to the DAs, they
also contain three auxiliary parameters, namely, Borel mass parameter,
continuum threshold $s_0$, and the parameter $\beta$ in the interpolating 
current. Since any physically measurable quantity should be independent of
them, we need to look for regions of $M^2$, $s_0$, and in which the results
of the vector meson--octet baryon coupling constants are practically
independent of these parameters.

The upper bound for the Borel parameter $M^2$ is determined by demanding
that the higher states and continuum contributions to a correlation
function should be less than half the value of the same correlation
function. The lower bound of $M^2$ can be determined by requiring that the
contribution of the highest term with the power of $1/M^2$ be less than
25\%. Using these restrictions, we obtain the working region for the Borel
parameters. The continuum threshold is varied in the regions
$s_0=(m_B+0.5)^2$ and $s_0=(m_B+0.7)^2$.

To demonstrate the analysis, in Figs. 1 and 2, we depict the dependence of 
$f_1^{p \rar p \rho_0}$ and
$f_1^{p \rar p \rho_0}+f_2^{p \rar p \rho_0}$ on $M^2$ at three different
values of the parameter $\beta$, and at two fixed values of $s_0$. The results
for $f_1^{p \rar p \rho_0}$ and $f_1^{p \rar p \rho_0}+f_2^{p \rar p \rho_0}$ 
exhibited by these figures show good stability with respect to the variation
of $M^2$ in its working domain. As has already been noted, the sum rules
contain another arbitrary parameter $\beta$, and with similar reasoning, we
should find such region of $\beta$, in which the results for the coupling
constants are independent of it. For this purpose, in Figs. 3 and 4, we
present the dependence  $f_1^{p \rar p \rho_0}$ and $f_1^{p \rar p
\rho_0}+f_2^{p \rar p \rho_0}$ on $\cos\theta$, where $\theta$ is defined as
$\tan\theta=\beta$. From these figures one can conclude that, the working region
for the unphysical parameter $\beta$ is $-0.5<\cos\theta<0.3$ for $f_1^{p \rar p
\rho_0}$, and $-0.7<\cos\theta<0.1$ for $f_1^{p \rar p \rho_0}+f_2^{p \rar p
\rho_0}$, where the coupling constants $f_1^{p \rar p \rho_0}$ and $f_1^{p
\rar p \rho_0}+f_2^{p \rar p \rho_0}$ are insensitive to the variation of
$\beta$. As a result of these considerations, we find that $f_1^{p \rar p
\rho_0}=-2.9 \pm 0.9$ and $f_2^{p \rar p \rho_0}=19.7 \pm 2.8$. 

Performing similar analysis, the results for the other coupling constants 
of vector mesons with octet baryons are presented in Table 1. For
completeness, we also present the existing results in literature in the same
Table. Note that in this Table we present only those results which are not
obtained from each other by a simple $SU(2)$ and isotopic spin relations.
We also would like to remind that, the signs of the residues are not
fixed by the sum rules. this leaves an ambiguity in the signs of any seven
(since an overall sign does not effect the coupling) couplings. These signs
have already been fixed in \cite{Roc07} to follow the $SU(3)_f$ symmetry.
In this work we follow the same sign convention. The error
bars in the table take into account only the uncertainties due to the
variations of the auxiliary parameters and the uncertainties in the input
parameters.

% .........................................................

\begin{table}[h]

\renewcommand{\arraystretch}{1.3}
\addtolength{\arraycolsep}{-0.5pt}
\small
$$
\begin{array}{|l|r@{\pm}l|r@{.}l|r@{\pm}l|r@{.}l|c|c|c|}
\hline \hline  
 \multirow{2}{*}{$f_1^{\mbox{\small{\,channel}}}$}             &  
 \multicolumn{4}{c|}{\mbox{General current}}   & 
 \multicolumn{4}{c|}{\mbox{Ioffe current}} & 
 \multirow{2}{*}{\mbox{QSR~\cite{Roc04}}}  & 
 \multirow{2}{*}{\mbox{QSR~\cite{Roc05}}}          & \multirow{2}{*}{\mbox{QSR~\cite{Roc06}}} \\ 
					    &    \multicolumn{2}{c}{\mbox{Result}}  & 
					\multicolumn{2}{c|}{\mbox{$SU(3)_f$}} &    \multicolumn{2}{c}{\mbox{Result}}   & 
                                        \multicolumn{2}{c|}{\mbox{$SU(3)_f$}} & & & \\ \hline
 f_1^{p \rar p \rho^0}                      & -2.5&1.1  & - 1&7 & -  5.9&1.3  & - 6&4   & 2.5 \pm 0.2    & 2.4 \pm 0.6 &  3.2 \pm 0.9 \\ 
 f_1^{p \rar p \omega}                      & -8.9&1.5  & -10&3 & -  8.2&0.4  & - 9&6   & 18 \pm 8~~     & 7.2 \pm 1.8 & \mbox{---}   \\ 
 f_1^{\Xi^0 \rar \Xi^0 \rho^0}              & -4.2&2.1  & - 4&3 & -  2.0&0.2  & - 1&6   & \mbox{---}     & 2.4 \pm 0.6 & 1.5 \pm 1.1   \\ 
 f_1^{\Sigma^0 \rar \Lambda \rho^0}         &  1.9&0.7  &   1&5 & -  3.0&0.5  & - 2&8   & \mbox{---}     & \mbox{---}  & \mbox{---}   \\ 
 f_1^{\Lambda \rar \Sigma^+ \rho^-}         &  1.9&0.7  &   1&5 &  - 2.8&0.6  & - 2&8   & \mbox{---}     & \mbox{---}  & \mbox{---}   \\ 
 f_1^{\Sigma^+ \rar \Sigma^0 \rho^+}        &  7.2&1.2  &   6&0 &    8.5&0.8  &   8&0   & \mbox{---}     & \mbox{---}  & \mbox{---}   \\ 
 f_1^{\Sigma^+ \rar \Lambda \rho^+}         &  2.0&0.6  &   1&5 &-   2.8&0.6  & - 2&8   & \mbox{---}     & \mbox{---}  & \mbox{---}   \\ 
 f_1^{p \rar \Lambda K^{\ast +}}            &  5.1&1.8  &   4&4 &    7.4&0.8  &   8&3   & \mbox{---}     & \mbox{---}  & \mbox{---}   \\ 
 f_1^{\Sigma^- \rar n K^{\ast -}}           &  6.6&1.8  &   6&1 &    1.7&0.4  &   2&3   & \mbox{---}     & \mbox{---}  & \mbox{---}   \\ 
 f_1^{\Xi^0 \rar \Sigma^+ K^{\ast -}}       & -2.3&1.7  & - 2&4 & - 10.0&1.8  & - 9&1   & \mbox{---}     & \mbox{---}  & \mbox{---}   \\ 
 f_1^{\Xi^- \rar \Lambda K^{\ast -}}        & -5.9&0.7  & - 5&8 & -  6.2&0.4  & - 5&5   & \mbox{---}     & \mbox{---}  & \mbox{---}   \\ 
 f_1^{\Sigma^0 \rar \Xi^0 K^{\ast 0}}       &  1.6&1.0  &   1&7 &    7.1&1.3  &   6&4   & \mbox{---}     & \mbox{---}  & \mbox{---}   \\ 
 f_1^{\Lambda \rar \Xi^0 K^{\ast 0}}        & -6.0&0.7  & - 5&9 & -  6.2&0.2  & - 5&5   & \mbox{---}     & \mbox{---}  & \mbox{---}   \\ 
 f_1^{n \rar \Sigma^0 K^{\ast 0}}           & -4.0&0.7  & - 4&3 & -  1.5&0.3  & - 1&6   & \mbox{---}     & \mbox{---}  & \mbox{---}   \\ 
 f_1^{\Lambda \rar \Lambda \omega}          & -7.1&1.1  & - 7&7 & -  4.8&0.2  & - 4&8   & \mbox{---}     & 4.8 \pm 1.2 & \mbox{---}   \\ 
 f_1^{\Xi^0 \rar \Xi^0 \phi}                & -9.5&2.5  & - 8&5 & - 13.5&1.6  & -11&3   & \mbox{---}     & \mbox{---}  & \mbox{---}   \\ 
 f_1^{\Lambda \rar \Lambda \phi}            & -5.3&1.5  & - 3&6 & -  8.0&1.0  & - 6&8   & \mbox{---}     & \mbox{---}  & \mbox{---}   \\ 
 f_1^{\Sigma^0 \rar \Sigma^0 \phi}          & -6.0&0.8  & - 6&1 & - 0.25&0.50 & - 2&3   & \mbox{---}     & \mbox{---}  & \mbox{---}   \\
\hline \hline
\end{array}
$$
\caption{The values of the electric coupling constants for various channels.}
\renewcommand{\arraystretch}{1}
\addtolength{\arraycolsep}{-1.0pt}

\end{table}

\begin{table}[h]

\renewcommand{\arraystretch}{1.3}
\addtolength{\arraycolsep}{-0.5pt}
\small
$$
\begin{array}{|l|r@{\pm}l|r@{.}l|r@{\pm}l|r@{.}l|c|c|c|}
\hline \hline  
 \multirow{2}{*}{$(f_1+f_2)^{\mbox{\small{\,channel}}}$}             &  
 \multicolumn{4}{c|}{\mbox{General current}}   & 
 \multicolumn{4}{c|}{\mbox{Ioffe current}} & 
 \multirow{2}{*}{\mbox{QSR~\cite{Roc04}}}  & 
 \multirow{2}{*}{\mbox{QSR~\cite{Roc05}}}          & \multirow{2}{*}{\mbox{QSR~\cite{Roc06}}} \\ 
					    &    \multicolumn{2}{c}{\mbox{Result}}  & 
					\multicolumn{2}{c|}{\mbox{$SU(3)_f$}} &    \multicolumn{2}{c}{\mbox{Result}}  & 
                                        \multicolumn{2}{c|}{\mbox{$SU(3)_f$}} & & & \\ \hline
 (f_1+f_2)^{p \rar p \rho^0}                      &  19.7&2.8  &  21&4 &   22.7&1.3   &  24&7   & 21.6 \pm 6.6   &  10.1  \pm 3.7 &  36.8 \pm 13  \\ 
 (f_1+f_2)^{p \rar p \omega}                      &  14.5&2.6  &  15&0 &   21.2&1.2   &  25&7   & 32.4 \pm 14.4  &~\, 5.0 \pm 1.2 & \mbox{---}    \\ 
 (f_1+f_2)^{\Xi^0 \rar \Xi^0 \rho^0}              & - 2.8&1.6  & - 3&2 & -  0.24&0.24 &   0&5   & \mbox{---}     &\! -3.6 \pm 1.6 & - 5.3 \pm 3.3 \\ 
 (f_1+f_2)^{\Sigma^0 \rar \Lambda \rho^0}         &  13.8&2.7  &  14&2 &   15.1&0.9   &  14&0   & \mbox{---}     & \mbox{---}     & \mbox{---}    \\ 
 (f_1+f_2)^{\Lambda \rar \Sigma^+ \rho^-}         &  14.3&2.9  &  14&2 &   15.1&0.8   &  14&0   & \mbox{---}     & \mbox{---}     & \mbox{---}    \\ 
 (f_1+f_2)^{\Sigma^+ \rar \Sigma^0 \rho^+}        & -17.8&2.2  & -18&2 & - 27.9&1.8   & -25&2   & \mbox{---}     &~\, 7.1 \pm 1.0 & 53.5 \pm 19   \\ 
 (f_1+f_2)^{\Sigma^+ \rar \Lambda \rho^+}         &  14.3&2.9  &  14&2 &   15.1&0.8   &  14&0   & \mbox{---}     & \mbox{---}     & \mbox{---}    \\ 
 (f_1+f_2)^{p \rar \Lambda K^{\ast +}}            & -22.9&4.2  & -22&9 & - 27.3&1.5   & -28&8   & \mbox{---}     & \mbox{---}     & \mbox{---}    \\ 
 (f_1+f_2)^{\Sigma^- \rar n K^{\ast -}}           &   3.8&2.8  &   4&5 & -  0.79&0.05 & - 0&7   & \mbox{---}     & \mbox{---}     & \mbox{---}    \\ 
 (f_1+f_2)^{\Xi^0 \rar \Sigma^+ K^{\ast -}}       &  33.8&4.9  &  30&3 &   41.3&2.4   &  34&9   & \mbox{---}     & \mbox{---}     & \mbox{---}    \\ 
 (f_1+f_2)^{\Xi^- \rar \Lambda K^{\ast -}}        &  11.6&2.9  &   8&7 &   17.9&1.0   &  14&8   & \mbox{---}     & \mbox{---}     & \mbox{---}    \\ 
 (f_1+f_2)^{\Sigma^0 \rar \Xi^0 K^{\ast 0}}       & -24.6&4.8  & -21&4 & - 29.2&1.7   & -24&7   & \mbox{---}     & \mbox{---}     & \mbox{---}    \\ 
 (f_1+f_2)^{\Lambda \rar \Xi^0 K^{\ast 0}}        &  11.1&2.6  &   8&7 &   15.0&1.0   &  14&8   & \mbox{---}     & \mbox{---}     & \mbox{---}    \\ 
 (f_1+f_2)^{n \rar \Sigma^0 K^{\ast 0}}           & - 2.8&1.8  & - 3&2 &    0.56&0.04 &   0&5   & \mbox{---}     & \mbox{---}     & \mbox{---}    \\ 
 (f_1+f_2)^{\Lambda \rar \Lambda \omega}          &   1.6&0.6  &   1&8 &    7.1&0.5   &   9&1   & \mbox{---}     & - 5.7 \pm 1.0  & \mbox{---}    \\ 
 (f_1+f_2)^{\Xi^0 \rar \Xi^0 \phi}                &  22.8&6.4  &  25&7 &   37.7&2.5   &  35&6   & \mbox{---}     & \mbox{---}     & \mbox{---}    \\ 
 (f_1+f_2)^{\Lambda \rar \Lambda \phi}            &  19.3&5.0  &  18&7 &   22.0&1.4   &  23&5   & \mbox{---}     & \mbox{---}     & \mbox{---}    \\ 
 (f_1+f_2)^{\Sigma^0 \rar \Sigma^0 \phi}          & - 3.5&2.5  &   4&5 &   0.81&0.05  &   0&7   & \mbox{---}     & \mbox{---}     & \mbox{---}    \\
\hline \hline
\end{array}
$$
\caption{The values of the magnetic coupling constants for various channels.}
\renewcommand{\arraystretch}{1}
\addtolength{\arraycolsep}{-1.0pt}

\end{table}

% .........................................................

From the results summarized in Table (1), we can comment as follows:

a) For the coupling $f_1$: 
A comparison of the predictions on the coupling constants which are 
obtained using the most general form of the currents, with the ones obtained 
using the Ioffe current shows substantial difference for many
channels. For example, the coupling constants $f_1$
for the $p \rar \Sigma^+ K^{\ast 0}$, $ \Xi^0 \rar \Sigma^0 \bar{K}^{\ast 0}$, 
$\Sigma^- \rar n K^{\ast -}$, $n \rar p \rho^-$, $\Xi^0 \rar \Sigma^- K^{\ast
-}$, $\Sigma^0 \rar \Xi^0 K^{\ast 0}$ channels for the general
case, differ considerably from the prediction of the Ioffe current.
Especially, the difference between the predictions of the above--mentioned
currents for the $\Sigma^0 \rar \Sigma^0 \phi$ transition is worth
mentioning. While the Ioffe current predicts $f_1 \simeq 0$, the general
current case predicts $f_1 \simeq - 5$. The sign of $f_1$ for the 
$\Sigma^0 \rar \Lambda \rho^0$, $\Sigma^+ \rar \Lambda \rho^+$, 
transitions differ from those predicted by the Ioffe current.
Our predictions on the coupling constant $f_1$ for the $p \rar p \rho^0$ 
within their error limits, are closer to the results predicted by
\cite{Roc04}, \cite{Roc05} and \cite{Roc06}. There is considerable 
difference for the $p \rar p \omega$ transition between our result
compared to that obtained in \cite{Roc04}, but our result is close to the
results of \cite{Roc05}.

b) For the coupling $f_1+f_2$: 
Except the $\Xi^0 \to \Xi^0 \rho^0$, $\Sigma^+ \rar \Sigma^0 \rho^+$,
 $\Sigma^- \rar n K^{\ast -}$, $\Xi^0 \to \Xi^0 \phi$,
$n \rar \Sigma^0 K^{\ast 0}$, $p \rar \Sigma^+ K^{\ast 0}$,
$\Lambda \rar \Lambda \omega$ and $\Sigma^0 \rar \Sigma^0 \phi$ transitions,
our predictions for the general current is in good agreement with the
predictions of Ioffe current.

These discrepancies between the coupling constants obtained using the general
form of the baryon current and the Ioffe current can be explained as
follows. For many channels the value $\beta=-1$ lies outside the stability
region of $\beta$, as a result of which considerable differences appear between the
predictions of the above--mentioned baryon currents, making the predictions
less reliable.

In the Tables, we have also presented in the columns labeled $SU(3)_f$
the best fits to our results of the
$SU(3)_f$ expressions given in Eq. (\ref{eoc01}). The $SU(3)_f$ fits in
Table--1 corresponds to the central values of ${\cal F}=-3.0 \pm 0.5$,
${\cal D}=1.3 \pm 0.6$ and ${\cal F}=-4.2 \pm 0.7$,
${\cal D}=-2.7 \pm 1.0$ for the general and Ioffe current, respectively. For
the central values, these yield $\alpha_E=1.6$ and $\alpha_E=0.61$,
respectively, both of which deviates from VDM model prediction $\alpha_E = 1$
considerably.

In Table--2, the $SU(3)_f$ fit value corresponds to ${\cal F}=9.2 \pm 1.0$,
${\cal D}=12.4 \pm 1.4$ and ${\cal F}=12.7 \pm 1.8$, ${\cal D}=12.2 \pm 1.8$
for the general form of the baryon current and 
$\beta=-1$ baryon current, respectively. For the $\alpha_M$ value of
the magnetic type coupling, these predictions yield $\alpha_M=0.43$ and
$\alpha_M=0.85$, respectively. $\alpha_M$ is also calculated in \cite{Roc18}
using the soft core potential and it is predicted to be $\alpha_M=0.44$,  
in agreement with the prediction of the general current.
           
In conclusion, the strong coupling constants of the vector mesons with octet
baryons are investigated in LCSR. It is proven that all coupling constants
can be written in terms of three universal functions, which at exact
$SU(3)_f$ symmetry case reduces to ${\cal F}$ and ${\cal D}$ couplings. The
numerical values of the electric and magnetic couplings  are obtained.

\section*{Acknowledgments}

V. S. Zamiralov is grateful to TUBITAK for the fund provided during his
visit to METU. V. S. Z. also likes to thank Physics Department of METU for
their hospitality. A. O. would like to thank TUBITAK for the fund provided
through the project 106T333. 

\newpage

\bAPP{A}{}

In this appendix we present the explicit expressions of the six Borel transformed invariant
functions.

\subsection*{Electric Type Coupling}

The electric type coupling is determined by the coefficient of the structure $\not\!p \not\!\varepsilon \not\!q$: 
\baeeq
\label{eocap01}
&& e^{m_V^2/4M^2} \Pi_1^{f_1}(u,d,s) = - {1\over 96\pi^2} M^4 (1+\beta^2) f_V^\parallel m_V 
\phi_V^\parallel(u_0) \nnb \\
\ar {1\over 384 M^2 \pi^2} (1-\beta) f_V^\perp m_V^2   
\Bigg(\gamma_E - \ln {M^2\over \Lambda^2}\Bigg) \Big\{ 
\GG [- m_d (1-\beta) \nnb \\
\ar 3 m_s (1+\beta) ] \psi_{3;V}^\parallel(u_0) +
4 M^2 m_V^2 \Big( [- m_d (1-\beta) - 9 m_s (1+\beta) ] i_1({\cal T},1) \nnb \\
\ar 4 m_s (1+\beta) i_1({\cal T}_3+ 2 {\cal T}_4,1) \Big) \Big\} \nnb \\
\ar {1\over 192 M^8}(1-\beta^2) f_V^\parallel m_V^3 \Big[ \GG m_0^2 (m_s \dd + m_d \sp)
\widetilde{\widetilde{i}}_4(\Bbb{C}) \Big] \nnb \\
\ar {1\over 288 M^6} (1-\beta^2) f_V^\parallel m_V^3 \Big[ 3 \GG (m_s \dd +    
m_d \sp) \widetilde{\widetilde{i}}_4(\Bbb{C}) \nnb \\
\ar 4 m_0^2 m_V^2 (m_s \dd - m_d \sp )
i_0(\widetilde{\Psi},1) \Big] \nnb \\
\ar {1\over 1152 M^4 \pi^2} m_V^4 f_V^\perp \GG (1-\beta) [m_d
(1-\beta) + m_s (1+\beta)] i_1({\cal T}-{\cal T}_4,1) \nnb \\
\ar {1\over 72 M^4} m_V^3 f_V^\parallel (1-\beta^2) \Big[
3 m_0^2 (m_s \dd + m_d \sp) \widetilde{\widetilde{i}}_4(\Bbb{C}) +
4 m_V^2 (m_s \dd - m_d \sp) i_0(\widetilde{\Psi},1) \Big] \nnb \\
\ar {1\over 288 M^2} m_V^3 f_V^\parallel \Big\{
2 [ m_s \sp (1+\beta)^2 - m_d \dd (3 \beta^2 + 2 \beta + 3) ] \Big[ i_2({\cal A},1) 
- i_2({\cal V},1-2 v) \Big] \nnb \\
\ek 8 [ m_s \sp \beta -  m_d \dd (2 \beta^2 +\beta +2)] i_2(\Phi,1-2 v)
+ 4 [ m_s \sp (1+\beta^2) \nnb \\
\ek 2 m_d \dd (2 \beta^2 +\beta +2)] i_2(\widetilde{\Phi},1) 
+ 4 ( m_d \dd +  m_s \sp) (1+\beta^2) i_2(\Psi,1-2 v) \nnb \\
\ek 4 [2 m_s \sp \beta +  m_d \dd (1+\beta^2)] 
i_2(\widetilde{\Psi},1)
%\ar {1\over 432 M^2} m_0^2 m_V^2 f_V^\perp (1-2 v) \widetilde{i}_4(\Bbb{B}_T)
%[3 \dd (1-\beta)^2 -  5 \sp (1-\beta^2)] \nnb \\
%
+ 12 [ 6 (1-\beta^2) (m_d \sp + m_s \dd) \nnb \\
\ar (m_d \dd + m_s \dd) (5 \beta^2 + 2 \beta + 5) ] \widetilde{\widetilde{i}}_4(\Bbb{C})
%\ar {1\over 864 M^2} m_0^2 m_V^2 f_V^\perp \widetilde{i}_4(\Bbb{C}_T) (1-2 v)   
%[ 3 \dd (1-\beta)^2 - 5 \sp (1-\beta^2) ] \nnb \\
+ 3 (m_d \dd + m_s \sp)
(1+\beta^2) \Bbb{A}(u_0) \Big\} \nnb \\
\ar {1\over 432 M^2} m_0^2 m_V f_V^\parallel  (m_d \dd + m_s
\sp) (3 \beta^2 + 2 \beta +3) \phi_V^\parallel(u_0) \nnb \\
\ar {1\over 36 M^2} m_V^4 f_V^\perp  (1-\beta) [ \dd (1-\beta) +
\sp (1+\beta)] i_1({\cal T}-{\cal T}_4,1) \nnb \\
\ar {1\over 576 M^2\pi^2} m_V^2 f_V^\perp 
\Big\{ \GG [m_d (1-\beta)^2 - 3 m_s (1-\beta^2)] \nnb \\
\ek 8 m_0^2 \pi^2 [\dd (1-\beta)^2 - 4
\sp (1-\beta^2)]\Big\} \psi_{3;V}^\parallel(u_0) \nnb \\
\ar {1\over 384 \pi^2} m_V^2 M^2 \Big\{ (1+\beta^2) f_V^\parallel \Big( 3 m_V \Bbb{A}(u_0)
- 4 m_V i_2({\cal A},1) + 4 m_V [ i_2(4 \Phi + 2 \Psi + {\cal V},1-2 v) \Big) \nnb \\
\ek 4 m_V f_V^\parallel \Big[ (3 \beta^2 + 2\beta + 3) i_2(\widetilde{\Phi},1) + 
(1+\beta)^2 i_2(\widetilde{\Psi},1)
- 3 (5 \beta^2 + 2 \beta + 5) \widetilde{\widetilde{i}}_4(\Bbb{C}) \Big] \nnb \\
\ek 12 f_V^\perp  (1-\beta) \Big[ m_d (1-\beta) - 3 m_s (1+\beta) \Big]
\psi_{3;V}^\parallel(u_0) \Big\} \nnb \\
\ek {1\over 24} m_V \Big\{ f_V^\parallel  (m_d \dd + m_s \sp)
(1+\beta^2) \phi_V^\parallel(u_0) \nnb \\
\ek 2 m_V f_V^\perp  (1-\beta) [ \dd (1-\beta) - 3 \sp
(1+\beta) ] \psi_{3;V}^\parallel(u_0) \Big\} \nnb \\
\ek {1\over 12 \pi^2} (1+\beta^2)  m_V^5 f_V^\parallel\Big[ 
i_0(\Psi,1-2 v) - i_0(\widetilde{\Psi},1)\Big] \nnb \\
\ek {1\over 96 \pi^2} m_V^4 f_V^\perp (1-\beta) \Big\{ [ 3 m_d (1-\beta) +
11 m_s (1+\beta) ]  i_1({\cal T},1) - 4 m_s (1+\beta) i_1({\cal T}_3,1) \nnb \\
\ek 2 [ m_d (1-\beta) + 5 m_s (1+\beta) ] i_1({\cal T}_4,1) \Big\}~, \\ \nnb \\
\label{??}
&& e^{m_V^2/4M^2} \Pi_2^{f_1}(u,d,s) =
- {1\over 96\pi^2} M^4 (3 \beta^2 + 2\beta +3) f_V^\parallel m_V 
\phi_V^\parallel(u_0) \nnb \\
\ek {1\over 384 M^2 \pi^2} (m_d + m_u) (1-\beta^2) f_V^\perp m_V^2   
\Bigg(\gamma_E - \ln {M^2\over \Lambda^2}\Bigg) \Big[  
4 M^2 m_V^2  i_1(9 {\cal T}-8 {\cal T}_3-4 {\cal T}_4,1) \nnb \\
\ek 3 \GG \psi_{3;V}^\parallel(u_0) \Big] \nnb \\
\ek {1\over 576 M^8} \GG f_V^\parallel m_0^2 m_V^3  (1-\beta)^2 (m_u \dd + m_d \uu) 
 \widetilde{\widetilde{i}}_4(\Bbb{C})  \nnb \\
\ek {1\over 288 M^6} f_V^\parallel m_V^3 (1-\beta)^2 (m_u \dd + m_d \uu)  
\Big[ \GG \widetilde{\widetilde{i}}_4(\Bbb{C})
+ 4 m_0^2 m_V^2 i_0(\widetilde{\Psi},1) \Big] \nnb \\
\ek {1\over 1152 M^4 \pi^2} m_V^4 f_V^\perp \GG (1-\beta^2) (m_u + m_d) 
i_1({\cal T}-{\cal T}_4,1) \nnb \\
\ek {1\over 18 M^4} m_V^5 f_V^\parallel (1-\beta)^2 (m_u \dd + m_d \uu)
i_0(\widetilde{\Psi},1) \nnb \\
\ek {1\over 288 M^2} m_V^3 f_V^\parallel 
(m_u \uu + m_d \dd) (3 \beta^2 + 2 \beta + 3) \Big[ 2 i_2({\cal A},1)
- 2 i_2(4 \Phi+2 \Psi + {\cal V},1-2 v) \nnb \\
\ek 3 \Bbb{A}(u_0) \Big]\nnb \\
\ek {1\over 72 M^2} m_V^3 f_V^\parallel \Big\{
3 [ 2 (1-\beta)^2 (m_u \dd + m_d \uu) - (m_u \uu + m_d \dd) (5 \beta^2 + 2 \beta + 5) ]
\widetilde{\widetilde{i}}_4(\Bbb{C}) \nnb \\
\ar 2 (1+\beta)^2 ( m_u \uu +  m_d \dd )  i_2(\widetilde{\Psi},1) \Big\} \nnb \\
%\ek {1\over 864 M^2} m_0^2 m_V^2 f_V^\perp \widetilde{i}_4(\Bbb{C}_T) (1-2 v)   
%(\uu + \dd ) (1-\beta^2) \nnb \\
\ar {1\over 216 M^2} m_V f_V^\parallel  (m_u \uu + m_d
\dd) (5 \beta^2 + 6 \beta +5) \Big[ m_0^2 \phi_V^\parallel(u_0)
- 3 m_V^2 i_2(\widetilde{\Phi},1) \Big] \nnb \\
\ek {1\over 36 M^2} m_V^4 f_V^\perp (\uu + \dd) [(1-\beta^2) i_1({\cal
T},1)-i_1({\cal T}_4,1)] \nnb \\
%\ek {1\over 432 M^2} m_0^2 m_V^2 f_V^\perp (1-2 v) \widetilde{i}_4(\Bbb{B}_T)
%(\uu + \dd) (1-\beta^2) \nnb \\
\ek {1\over 576 \pi^2 M^2} m_V^2 f_V^\perp  (1-\beta^2)
[ 3 \GG (m_u + m_d) - 40 m_0^2 \pi^2 (\uu + \dd) ] \psi_{3;V}^\parallel(u_0) \nnb \\
\ek {1\over 96 \pi^2} m_V^3 M^2 f_V^\parallel \Big\{ (3 \beta^2 + 2 \beta +3)
[i_2({\cal A},1) - 4 i_2(\Phi,1 - 2 v)] + 2 (5 \beta^2 + 6 \beta + 5)
i_2(\widetilde{\Phi},1) \nnb \\
\ek 2 (3 \beta^2 + 2 \beta +3) i_2(\Psi,1 - 2 v) - 4 (1 + \beta)^2
i_2(\widetilde{\Psi},1) -
(3 \beta^2 + 2 \beta +3)   i_2({\cal V},1 - 2 v) \Big\} \nnb \\
\ar {1\over 128 \pi^2}  m_V^2 M^2 \Big\{ f_V^\parallel m_V [
(3 \beta^2 + 2 \beta +3) \Bbb{A}(u_0) +
4 (5 \beta^2 + 2 \beta + 5) \widetilde{\widetilde{i}}_4(\Bbb{C}) ] \nnb \\
\ar 12 (m_u + m_d) (1-\beta^2) f_V^\perp \psi_{3;V}^\parallel(u_0) \Big\} \nnb \\
\ek {1\over 24} \Big[ m_V f_V^\parallel (3 \beta^2 + 2 \beta +3) 
(m_u \uu + m_d \dd) \phi_V^\parallel(u_0) + 
6 m_V^2 f_V^\perp (1-\beta^2)  (\uu + \dd) \psi_{3;V}^\parallel(u_0) \Big] \nnb \\
\ek {1\over 96 \pi^2} m_V^4 f_V^\perp 
(m_u + m_d) (1 - \beta^2 ) i_1(7{\cal T}-8{\cal T}_3-2 {\cal T}_4,1) \nnb \\
\ek {1\over 12 \pi^2} m_V^5 f_V^\parallel (3 \beta^2 + 2 \beta +3)
\Big[ i_0(\Psi,1-2 v) - i_0(\widetilde{\Psi},1) \Big]~, \\ \nnb \\
&& e^{m_V^2/4M^2} (\Pi_3^{f_1})^{asym}(u,d,s) =
-{1\over 768 \sqrt{2} M^2 \pi^2} f_V^\perp \GG m_V^2
(m_d-m_s) (1-\beta) (3+\beta) \nnb \\
\cp \Bigg(E_\gamma - \ln {M^2\over \Lambda^2}\Bigg)
[2 \widetilde{i}_4(\Bbb{B}_T) + \widetilde{i}_4(\Bbb{C}_T) ] \nnb \\
\ek {1\over 72 \sqrt{2} M^6} f_V^\parallel m_0^2 m_V^5 (m_d \sp - m_s \dd)
(1-\beta) (3+\beta) i_0(\Psi,1) \nnb \\
\ek {1\over 2304 \sqrt{2} M^4 \pi^2} f_V^\perp m_V^4 (m_d-m_s) \GG (1-\beta)
(3+\beta) i_1({\cal T} - 2 {\cal T}_4,1-2 v) \nnb \\
\ek {1\over 18 \sqrt{2} M^4}  f_V^\parallel m_V^5 (m_d \sp - m_s\dd) (1-\beta)
(3+\beta) i_0(\Psi,1) \nnb \\
\ar {1\over 3456 \sqrt{2} \pi^2 M^2} m_V^2 f_V^\perp (1-\beta)  
[ 3 \GG (m_d-m_s) (3+\beta) \nnb \\
\ek 16 m_0^2 \pi^2 (\dd -\sp) (5+2\beta) ]
[2 \widetilde{i}_4(\Bbb{B}_T) + \widetilde{i}_4(\Bbb{C}_T) ] \nnb \\
\ek {1\over 72 \sqrt{2} M^2} m_V^4 f_V^\perp (\dd-\sp) (1-\beta) (3+beta)
i_1({\cal T} - 2 {\cal T}_4,1-2 v) \nnb \\
\ar {1\over 48 \sqrt{2} M^2} f_V^\parallel m_V^3 (m_d \dd - m_s \sp)
(1+\beta)^2 \Big\{ 2 [i_2(\Phi,1) - i_2(\widetilde{\Phi},1-2 v)]  
- [i_2({\cal A},1-2 v) - i_2({\cal V},1)] \Big\} \nnb \\
\ek {1\over 64 \sqrt{2} \pi^2} m_V^2 M^2 (m_d-m_s) f_V^\perp (1-\beta)
(3+\beta) [2 \widetilde{i}_4(\Bbb{B}_T) + \widetilde{i}_4(\Bbb{C}_T) ] \nnb \\
\ar {1\over 24 \sqrt{2}} m_V^2 f_V^\perp (\dd-\sp) (1-\beta) (3+\beta)
[2 \widetilde{i}_4(\Bbb{B}_T) + \widetilde{i}_4(\Bbb{C}_T) ] \nnb \\
\ar {1\over 96 \sqrt{2} \pi^2} m_V^4 f_V^\perp (m_d-m_s) (1-\beta) (3+\beta)
i_1({\cal T} - 2 {\cal T}_4,1-2 v)~. 
\end{eqnarray}
 
\subsection*{Magnetic Type Coupling}
The magnetic type coupling is determined by the coefficient of the structure $(\varepsilon \cdot p) \not\!p$:
\begin{eqnarray} 
&& e^{m_V^2/4M^2} \Pi_1^{f_1+f_2}(u,d,s) =
{1\over 64 \pi^2} M^4 (1-\beta) f_V^\perp 
[m_d-m_s -(m_d+m_s) \beta] \phi_V^\perp(u_0) \nnb \\
\ar {1\over 192 \pi^2} M^4 m_V f_V^\parallel \Big\{ i_3({\cal V},1) + 
9 \psi_{3;V}^\perp(u_0) - \beta \Big[ 2 i_3({\cal A},1-2 v) \nnb \\
\ek \beta i_3({\cal V},1) - 3 (2 + 3 \beta) \psi_{3;V}^\perp(u_0) \Big] \Big\} \nnb \\
\ar {1\over 768 \pi^2} \Bigg\{ (1-\beta) f_V^\perp
\Bigg(\gamma_E - \ln {M^2\over \Lambda^2}\Bigg) 
\Big( 8 M^2 [m_d (1-\beta) + m_s (1+\beta)] m_V^2 i_2({\cal S},1) \nnb \\
\ar [m_d (1-\beta) - m_s (1+\beta)] (8 M^2 m_V^2 i_2(\widetilde{\cal S},1) 
+ \GG \phi_V^\perp(u_0)) \Big) \Big\} \nnb \\
\ar {1\over 384 M^6} m_V m_0^2 f_V^\parallel \GG (1-\beta^2) (m_s \dd +
m_d \sp)  \psi_{3;V}^\perp(u_0) \nnb \\
\ar {1\over 192 M^4} m_V f_V^\parallel \GG (1-\beta^2) (m_s \sp + m_d \dd)
\psi_{3;V}^\perp(u_0) \nnb \\
\ek {1\over 9216 M^2 \pi^2} m_V^2 f_V^\perp \GG (1-\beta) 
\Big\{4 [ m_d (1-\beta) + 3 m_s (1+\beta) ] 
i_2({\cal S},1)\nnb \\
\ar [ m_d (1-\beta) - m_s (1+\beta) ] 
[4 i_2({\cal T}_1-{\cal T}_2+{\cal T}_3-{\cal T}_4,1-2 v) - 3
\Bbb{A}_T(u_0)] \Big\}\nnb \\
%\ar {1\over 2304 M^2 \pi^2} m_V^2 f_V^\perp \GG
%(1-\beta) [ m_d (1-\beta) - m_s (1+\beta) ] i_2({\cal T}_2,1-2 v) \nnb \\
%\ek {1\over 2304 M^2 \pi^2} m_V^2 f_V^\perp  \GG
%(1-\beta) [ m_d (1-\beta) - m_s (1+\beta) ] i_2({\cal T}_3,1-2 v) \nnb \\
%\ar {1\over 2304 M^2 \pi^2} m_V^2 f_V^\perp \GG
%(1-\beta) [ m_d (1-\beta) - m_s (1+\beta) ] i_2({\cal T}_4,1-2 v) \nnb \\
%
\ar {1\over 144 M^2} m_V f_V^\parallel \Big\{m_V^2  
[ m_d \dd (1+\beta)^2 - m_s \sp (\beta^2 + 6 \beta + 1) ] 
i_2({\cal A},1-2 v)\nnb \\
\ek m_V^2  
[ m_d \dd (1+\beta)^2 - m_s \sp (3 \beta^2 + 2 \beta + 3) ] i_2({\cal V},1) \nnb \\
\ar 3 m_0^2 (1-\beta^2) ( m_d
\sp + m_s \dd) \psi_{3;V}^\perp(u_0) \Big\} \nnb \\
\ar {1\over 384 \pi^2} M^2 m_V^2 f_V^\perp (1-\beta) 
\Big\{ [m_d (1-\beta) - m_s (1+\beta)] \Big[ 4 i_2({\cal T}_1-{\cal T}_2
+{\cal T}_3-{\cal T}_4,1-2 v) \nnb \\
\ar 4 i_2(\widetilde{\cal S},1)
- 3 \Bbb{A}_T(u_0) \Big]
+ 8 [m_d (1-\beta) + 2 m_s (1+\beta)] i_2({\cal S},1) \Big\} \nnb \\
\ar {1\over 96 \pi^2} M^2 m_V^3 f_V^\parallel
\Big[ i_2({\cal V},1) - 2 \beta i_2({\cal A},1-2 v) + 
\beta^2 i_2({\cal V},1) \Big] \nnb \\
\ek {1\over 288} f_V^\perp (1-\beta) [\dd (1-\beta) - \sp (1+\beta)]
\Big[ 4 m_V^2 i_2({\cal T}_1-{\cal T}_2+{\cal T}_3-{\cal T}_4,1-2 v) \nnb \\
\ek 3 m_V^2 \Bbb{A}_T(u_0) + 12 M^2 \phi_V^\perp(u_0) \Big] \nnb \\
\ek {1\over 72} m_V^2 f_V^\perp (1-\beta) [\dd (1-\beta) + 3 \sp (1+\beta)]
i_2({\cal S},1) \nnb \\
\ek {1\over 1152 \pi^2} f_V^\perp \GG (1-\beta) [m_d (1-\beta) - m_s (1+\beta)] \phi_V^\perp(u_0) \nnb \\
\ar {1\over 432} m_0^2 f_V^\perp (1-\beta) [3 \dd (1-\beta) - 2 \sp (1+\beta)] \phi_V^\perp(u_0) \nnb \\
\ar {1\over 288} m_V f_V^\parallel \Big\{ [ m_d \dd (1+\beta)^2 
- m_s \sp (\beta^2+6 \beta+1) ] i_3({\cal A},1-2 v) \nnb\\
\ek [ m_d \dd (1+\beta)^2 
- m_s \sp (3 \beta^2+2 \beta+3) ] i_3({\cal V},1) \Big\} \nnb \\
\ar {1\over 48} m_V f_V^\parallel  \Big\{
\dd [6 m_s (1-\beta^2) + m_d (3 \beta^2 + 2 \beta + 3)] \nnb \\
\ar \sp [6 m_d (1-\beta^2) + m_s (3 \beta^2 + 2 \beta + 3)] \Big\} 
\psi_{3;V}^\perp(u_0)~, \\ \nnb \\
&& e^{m_V^2/4M^2} \Pi_2^{f_1+f_2}(u,d,s) =
{1\over 64 \pi^2} M^4 (1-\beta^2) f_V^\perp 
(m_u+m_d) \phi_V^\perp(u_0) \nnb \\
\ek {1\over 192 \pi^2} M^4 m_V f_V^\parallel \Big[ 
(\beta^2 + 6 \beta + 1) i_3({\cal A},1-2 v) - (3 \beta^2 + 2 \beta +3 ) i_3({\cal V},1) \nnb \\
\ar 3 (1+\beta)^2 \psi_{3;V}^\perp(u_0) \Big] \nnb \\
\ek {1\over 768 \pi^2} (m_u+m_d) f_V^\perp (1-\beta^2)
\Bigg(\gamma_E - \ln {M^2\over \Lambda^2}\Bigg)
\Big[ 8 M^2 m_V^2 i_2({\cal S}-\widetilde{\cal S},1) 
- \GG \phi_V^\perp(u_0) \Big] \nnb \\
\ar {1\over 1152 M^6} m_V  f_V^\parallel \GG (1-\beta)^2 (m_u \dd +
m_d \uu) (m_0^2 + 2 M^2) \psi_{3;V}^\perp(u_0) \nnb \\
\ar {1\over 9216 M^2 \pi^2} m_V^2 f_V^\perp  \GG (1-\beta^2) 
(m_u + m_d) \Big[ 3 \Bbb{A}_T(u_0) - 12 i_2({\cal S},1) \nnb \\
\ek 4 i_2({\cal T}_1-{\cal T}_2,+{\cal T}_3-{\cal T}_4,1-2 v) 
\Big] \nnb \\
%
%\ar {1\over 2304 M^2 \pi^2} m_V^2 f_V^\perp 
%\GG (1-\beta^2) (m_u + m_d) i_2({\cal T}_2,1-2 v) \nnb \\
%\ek {1\over 2304 M^2 \pi^2} m_V^2 f_V^\perp 
%\GG (1-\beta^2) (m_u + m_d) i_2({\cal T}_3,1-2 v) \nnb \\
%\ar {1\over 2304 M^2 \pi^2} m_V^2 f_V^\perp 
%\GG (1-\beta^2) (m_u + m_d) i_2({\cal T}_4,1-2 v) \nnb \\
\ek {1\over 144 M^2} m_V^3 f_V^\parallel 
(m_u \uu + m_d \dd) \Big[ (\beta^2+6 \beta+1) i_2({\cal A},1-2 v) 
- (3 \beta^2+2 \beta+3) i_2({\cal V},1) \Big] \nnb \\
\ar {1\over 384 \pi^2} M^2 m_V^2 f_V^\perp (1-\beta^2) (m_u + m_d)
\Big[ 4 i_2({\cal T}_1-{\cal T}_2+{\cal T}_3-{\cal T}_4,1-2 v) \nnb \\ 
\ar 4 i_2(2{\cal S}+\widetilde{\cal S},1) -
3 \Bbb{A}_T(u_0) \Big] \nnb \\
\ar {1\over 96 \pi^2} M^2 m_V^3 f_V^\parallel
\Big[(3 \beta^2+2 \beta+3) i_2({\cal V},1) - 
(\beta^2 + 6\beta +1) i_2({\cal A},1-2 v) \Big] \nnb \\
\ek {1\over 72} f_V^\perp (1-\beta^2) (\uu + \dd)
\Big\{ m_V^2 \Big[ i_2({\cal T}_1-{\cal T}_2+
{\cal T}_3-{\cal T}_4,1-2 v) \nnb \\
\ar 3 i_2({\cal S},1)\Big] + 3 M^2 \phi_V^\perp(u_0) \Big\} \nnb \\
\ar {1\over 864} f_V^\perp (1-\beta^2) (\uu+\dd) \Big[ 10 m_0^2 \phi_V^\perp(u_0) 
+ 9 m_V^2  \Bbb{A}_T(u_0) \Big] \nnb \\
\ek {1\over 1152 \pi^2} f_V^\perp \GG (1-\beta^2) (m_u+m_d) \phi_V^\perp(u_0) \nnb \\
\ek {1\over 288} m_V f_V^\parallel (m_u \uu + m_d \dd)
\Big[ (\beta^2+6 \beta+1) i_3({\cal A},1-2 v) - (3 \beta^2+2 \beta+3) i_3({\cal V},1) 
\Big] \nnb \\
\ek {1\over 48} m_V f_V^\parallel  \Big\{
\uu [m_u (1+\beta)^2 - 2 m_d (1- \beta)^2] -
\dd [2 m_u (1-\beta)^2 - m_d (1+\beta)^2] \Big\} \psi_{3;V}^\perp(u_0)~, \\ \nnb \\
&& e^{m_V^2/4M^2} ({\Pi_3^{f_1+f_2})^{asym}}(u,d,s) =
{1\over 96 \sqrt{2} \pi^2} (1-\beta) f_V^\perp m_V^2 M^2
(m_d-m_s) \Bigg(E_\gamma - \ln {M^2\over \Lambda^2}\Bigg) \nnb \\
\cp [ 4 (1+\beta) i_2({\cal T}_1-{\cal T}_2,1) +
(3+\beta) i_2({\cal T}_3-{\cal T}_4,1) ] \nnb \\
\ek {1\over 2304 \sqrt{2} \pi^2 M^2} m_V^2 (m_d-m_s)
f_V^\perp \GG (1-\beta) [ (1-\beta) i_2(\widetilde{\cal S},1-2 v)
+ (1+3 \beta) i_2({\cal T}_1-{\cal T}_2,1) ] \nnb \\
\ar {1\over 24 \sqrt{2} M^2} m_V (m_d \dd -m_s \sp) u_0 f_V^\parallel
q^2 (1+\beta)^2 [ i_2({\cal A},1) - i_2({\cal V},1-2 v) ] \nnb \\
\ar {1\over 96 \sqrt{2} \pi^2} m_V^2 (m_d-m_s) M^2 f_V^\perp (1-\beta)
[(1-\beta) i_2(\widetilde{\cal S},1-2 v) + (5+7 \beta)
i_2({\cal T}_1-{\cal T}_2,1) \nnb \\
\ar (3 +\beta) i_2({\cal T}_3-{\cal T}_4,1)] \nnb \\
\ar {1\over 96 \sqrt{2}} m_V (m_d \dd -m_s \sp) f_V^\parallel (1+\beta)^2
[i_3({\cal A},1) - i_3({\cal V},1-2 v) \nnb \\
\ek {1\over 72 \sqrt{2}} m_V^2 (\dd - \sp) f_V^\perp
(1-\beta) [(1-\beta) i_2(\widetilde{\cal S},1-2 v) +
(1+3 \beta) i_2({\cal T}_1-{\cal T}_2,1)]~.
%
%(* april 8, 2008 wednesday 15:00 *)
%
\eaeeq

\newpage

The functions $i_n$, $\widetilde{i}_4$ and
$\widetilde{\widetilde{i}}_4$ are defined as
\baeeq
\label{nolabel}
i_0(\phi,f(v)) \es \int {\cal D}\alpha_i \int_0^1 dv
\phi(\alpha_{\bar{q}},\alpha_q,\alpha_g) f(v) (k-u_0) \theta(k-u_0)~, \nnb \\
i_1(\phi,f(v)) \es \int {\cal D}\alpha_i \int_0^1 dv
\phi(\alpha_{\bar{q}},\alpha_q,\alpha_g) f(v) \theta(k-u_0)~, \nnb \\
i_2(\phi,f(v)) \es \int {\cal D}\alpha_i \int_0^1 dv
\phi(\alpha_{\bar{q}},\alpha_q,\alpha_g) f(v) \delta(k-u_0)~, \nnb \\
i_3(\phi,f(v)) \es \int {\cal D}\alpha_i \int_0^1 dv
\phi(\alpha_{\bar{q}},\alpha_q,\alpha_g) f(v) \delta^\prime(k-u_0)~, \nnb \\
\widetilde{i}_4(f(u)) \es \int_{u_0}^1 du f(u)~, \nnb \\
\widetilde{\widetilde{i}}_4(f(u)) \es \int_{u_0}^1 du (u-u_0) f(u)~, \nnb
\eaeeq
where 
\baeeq
k = \alpha_q + \alpha_g \bar{v}~,~~~~~u_0={M_1^2 \over M_1^2
+M_2^2}~,~~~~~M^2={M_1^2 M_2^2 \over M_1^2
+M_2^2}~.\nnb
\eaeeq

In the expressions for $\Pi_i^{(\alpha)}$, the contributions of the higher states and continuum are subtracted using the replacements:
\begin{eqnarray}
e^{-m_V^2/4 M^2} M^2 \left( \ln{M^2 \over \Lambda^2} - \gamma_E \right)  
		\rightarrow && \int_{m_V^2/4}^{s_0}  ds  e^{- s/M^2} \ln \frac{s-m_V^2/4}{\Lambda^2}\nonumber \\
e^{-m_V^2/4 M^2} \left( \ln{M^2 \over \Lambda^2} - \gamma_E \right) \rightarrow && \ln \frac{s_0 - m_V^2/4}{\Lambda^2} e^{-s_0/M^2} + 
	\frac{1}{M^2} \int_{m_V^2/4}^{s_0} ds e^{- s/M^2} \ln \frac{s-m_V^2/4}{\Lambda^2}\nonumber \\
e^{-m_V^2/4 M^2} \frac{1}{M^2} \left( \ln{M^2 \over \Lambda^2} - \gamma_E \right) \rightarrow && \frac{1}{M^2} \ln \frac{s_0 - m_V^2/4}{\Lambda^2} e^{-s_0/M^2}
+ \frac{1}{s_0-m_V^2/4}  e^{-s_0/M^2} 
\nonumber \\ &&
+  \frac{1}{M^4} \int_{m_V^2/4}^{s_0} ds e^{- s/M^2} \ln \frac{s-m_V^2/4}{\Lambda^2}\nonumber \\
e^{-m_V^2 / 4 M^2} M^{2n}\rightarrow && \frac{1}{\Gamma(n)} \int_{m_V^2/4}^{s_0} ds  e^{- s/M^2} \left( s - m_V^2/4 \right)^{n-1}\nonumber 
\end{eqnarray} 
\eAPP

\newpage

\section*{Figure captions}
{\bf Fig. (1)} The dependence of the electric coupling constant $f_1$ of
$p \rar p \rho^0$  transition on Borel mass $M^2$ for the three fixed values
of the parameter $\beta$: $\beta=-1,\pm 5$, and two fixed values of the vacuum threshold
$s_0$: $s_0=2.25~GeV^2$ and $s_0=2.75~GeV^2$ \\ \\ 
{\bf Fig. (2)} The dependence of the electric coupling constant $f_1$ of
$p \rar p \rho^0$  transition on $\cos\theta$ for the two fixed values of 
the vacuum threshold $s_0$: $s_0=2.25~GeV^2$ and $s_0=2.75~GeV^2$, and 
for the Borel mass at $M^2 = 1~GeV^2$. \\ \\
{\bf Fig. (3)} The same as in Fig. (1), but for the coupling constant
$(f_1+f_2)$. \\ \\
{\bf Fig. (4)} The same as in Fig. (2), but for the coupling constant
$(f_1+f_2)$.

\newpage

\begin{figure}
\vskip 3. cm
    \includegraphics{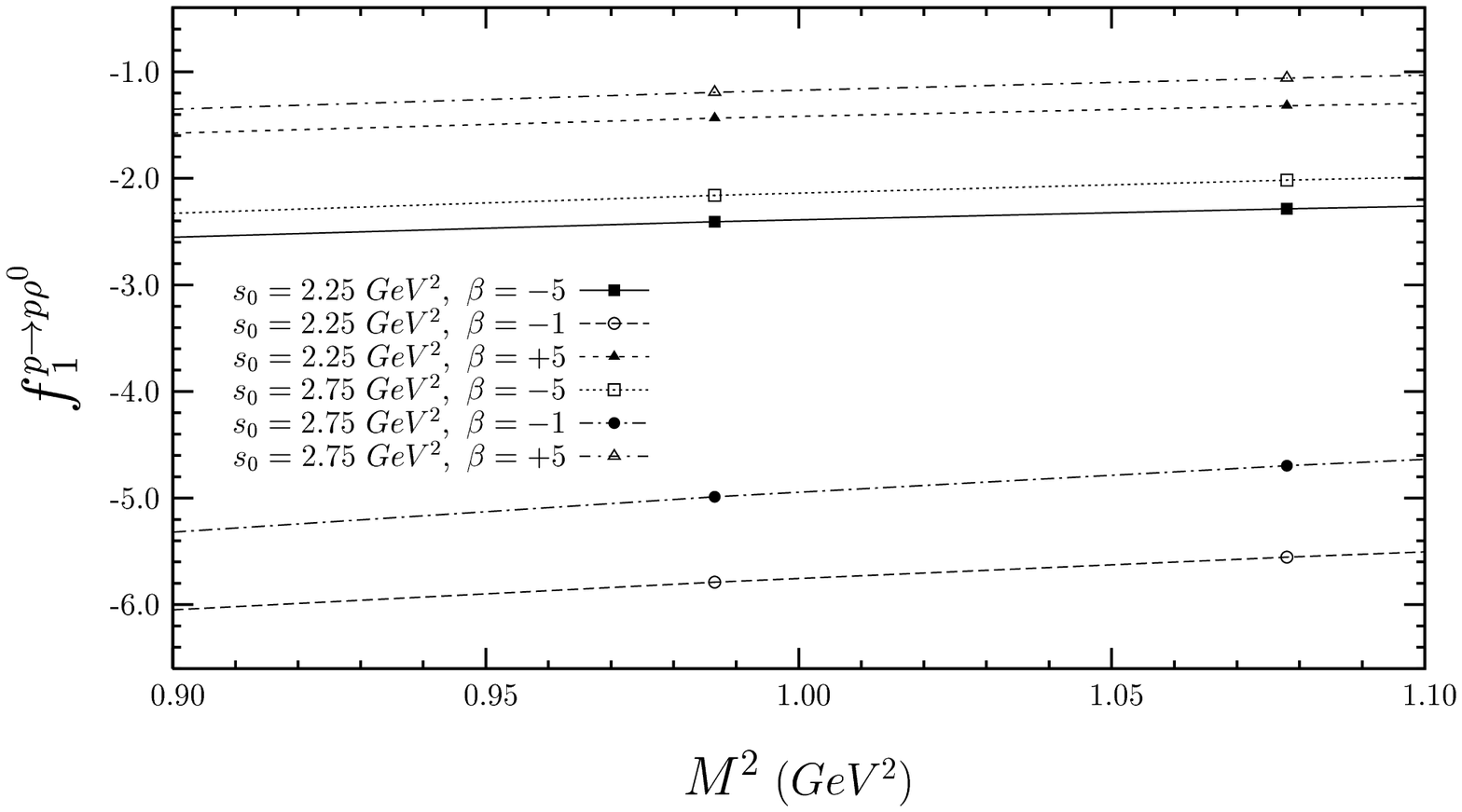}
\vskip 6.3cm
\caption{}
%\begin{center}
%{\bf Fig. 1--a}
%\end{center}
\end{figure}

\begin{figure}
\vskip 4.0 cm
    \includegraphics{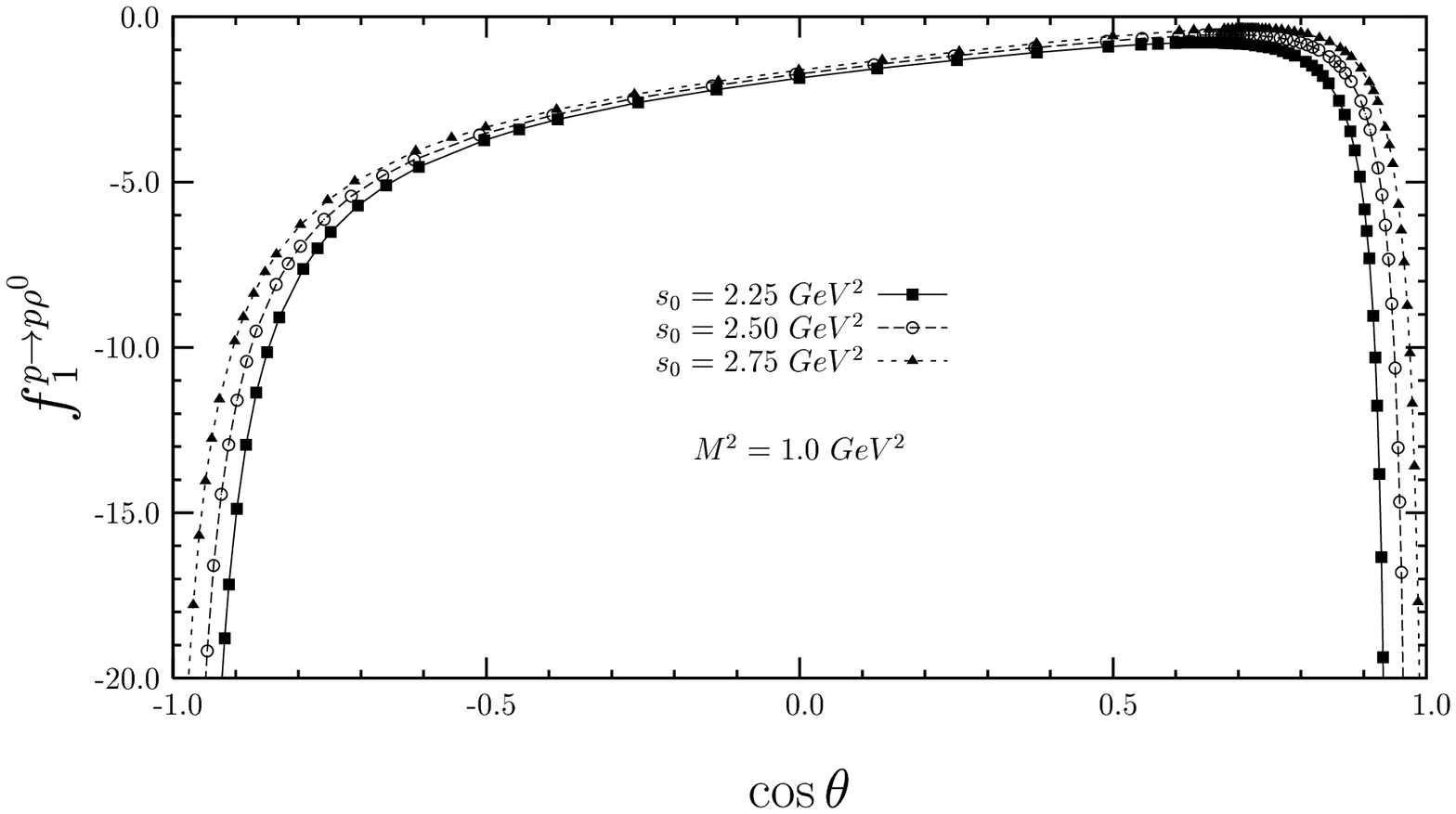}
\vskip 6.3 cm
\caption{}
%\begin{center}
%{\bf Fig. 1--b}
%\end{center}
\end{figure}

\newpage

\begin{figure}
\vskip 3. cm
    \includegraphics{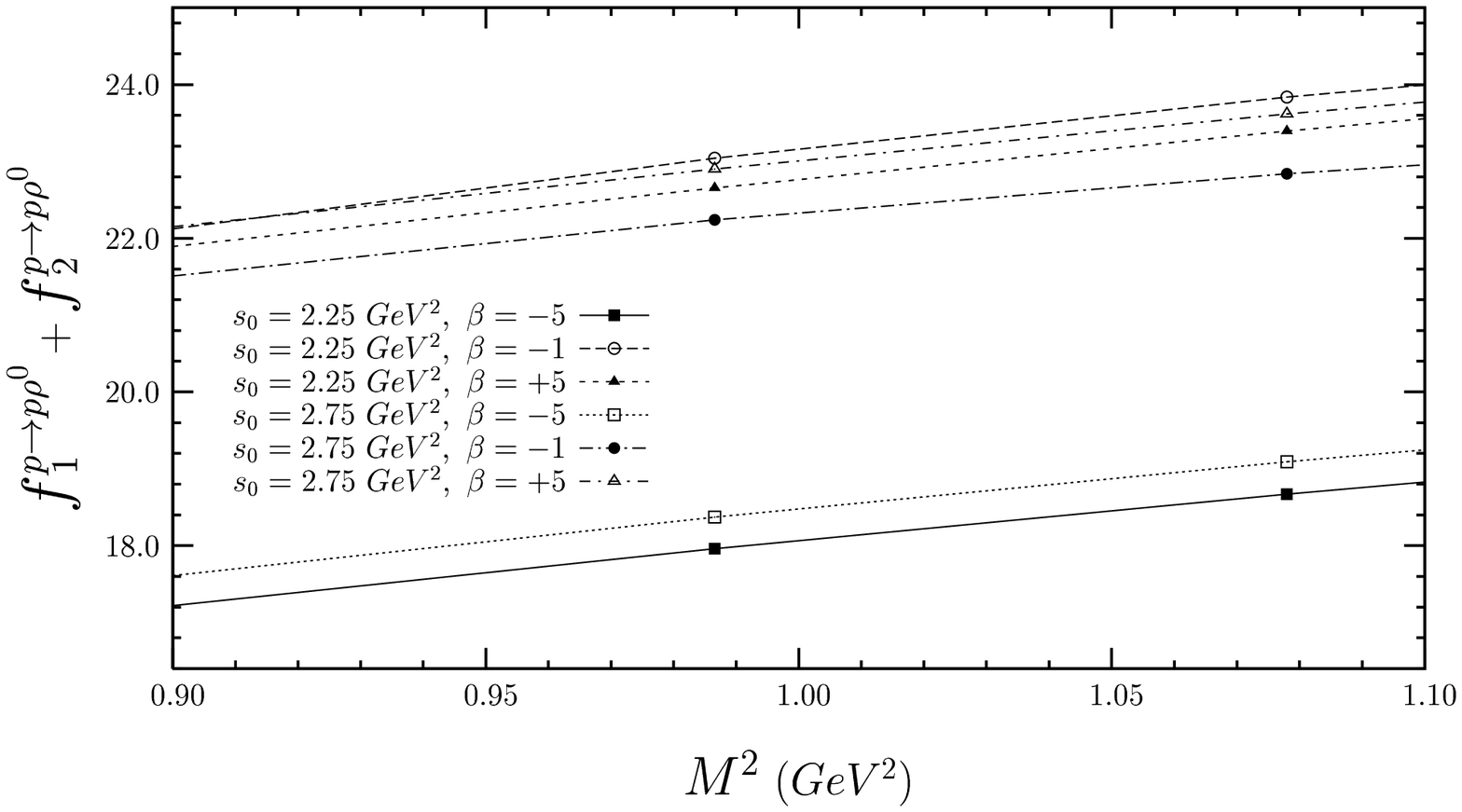}
\vskip 6.3cm
\caption{}
%\begin{center}
%{\bf Fig. 1--a}
%\end{center}
\end{figure}

\begin{figure}
\vskip 4.0 cm
    \includegraphics{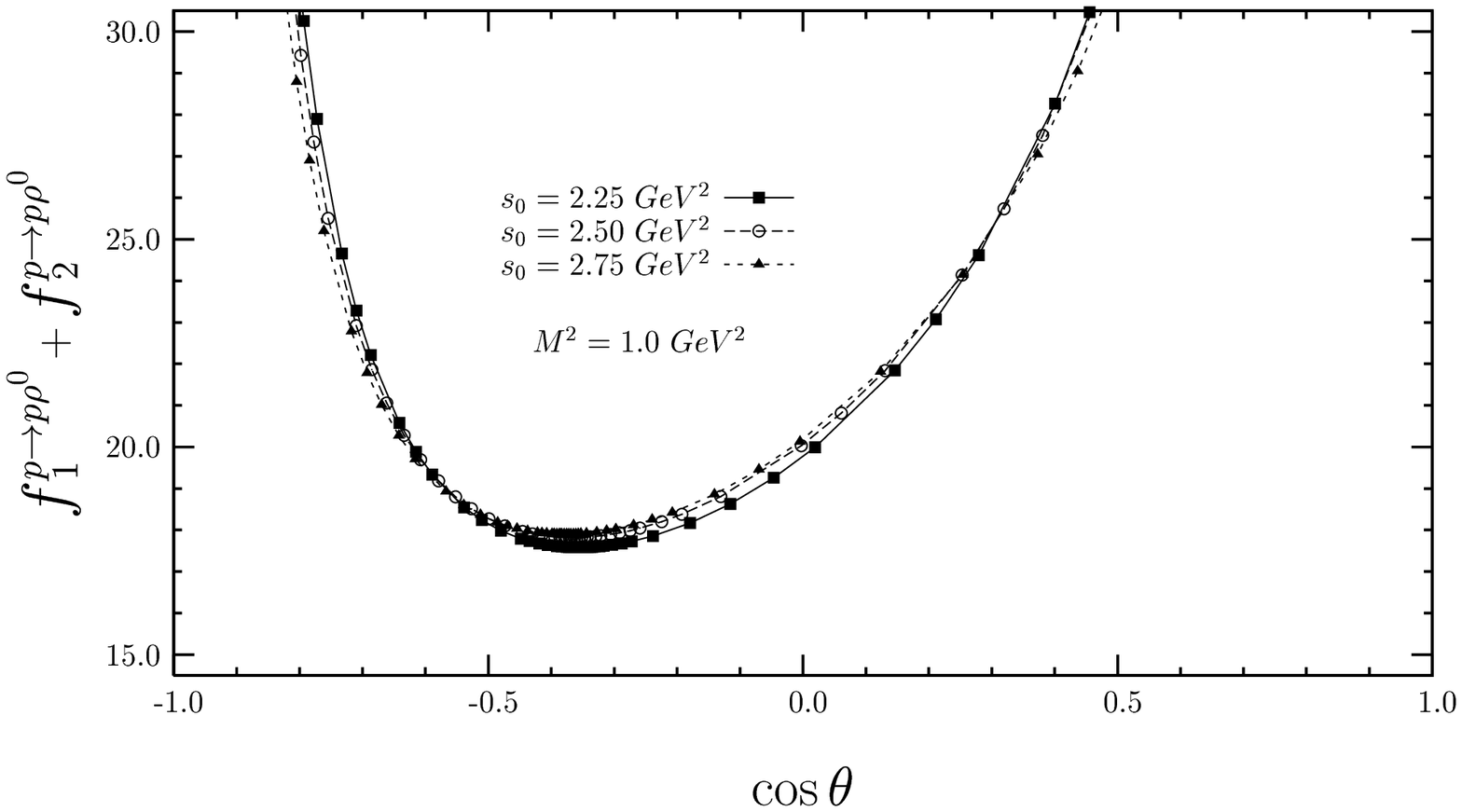}
\vskip 6.3 cm
\caption{}
%\begin{center}
%{\bf Fig. 1--b}
%\end{center}
\end{figure}

\end{document}